\documentclass[aps,prd,twocolumn,superscriptaddress,a4paper]{revtex4-2}

\usepackage[pdftex]{graphicx}
\usepackage{amsmath}
\usepackage{verbatim}
\usepackage{xcolor}
\usepackage{bm}
\usepackage{txfonts}
\usepackage{siunitx}
\usepackage{enumitem}
\usepackage[colorlinks=true,allcolors=blue]{hyperref}

\setcitestyle{super}

\makeatletter
\newcommand{\customlabel}[2]{%
   \protected@write \@auxout {}{\string \newlabel {#1}{{#2}{\thepage}{#2}{#1}{}} }%
   \hypertarget{#1}{}
}
\makeatother

\makeatletter
\renewcommand*{\fnum@figure}{{\normalfont\bfseries \figurename~\thefigure}}
\renewcommand*{\@caption@fignum@sep}{\textbf{~$\vert$~}}
\makeatother

\begin{document}
\renewcommand{\figurename}{\textbf{Fig.}}
\renewcommand{\thefigure}{\textbf{\arabic{figure}}}
\renewcommand{\thetable}{\textbf{\arabic{table}}}

\title{Measurement of the Casimir force between superconductors}

\author{Matthijs H.\ J.\ de Jong}
\affiliation{Department of Applied Physics, Aalto University, FI-00076 Aalto, Finland}

\author{Evren Korkmazgil}
\affiliation{Department of Applied Physics, Aalto University, FI-00076 Aalto, Finland}

\author{Louise Banniard}
\affiliation{Department of Applied Physics, Aalto University, FI-00076 Aalto, Finland}

\author{Mika A.\ Sillanpää}
\affiliation{Department of Applied Physics, Aalto University, FI-00076 Aalto, Finland}

\author{Laure Mercier de L{\'e}pinay} \email{laure.mercierdelepinay@aalto.fi}
\affiliation{Department of Applied Physics, Aalto University, FI-00076 Aalto, Finland}

\begin{abstract}
The Casimir force follows from quantum fluctuations of the electromagnetic field and yields a nonlinear attractive force between closely spaced conductive objects. 
Measuring the Casimir force in superconducting materials on either side of the transition should allow to isolate the specific contribution of low frequencies to the Casimir effect. 
There is significant interest in this contribution as it is suspected to be involved in an unexplained discrepancy between predictions and measurements of the Casimir force between normal metals. 
Here, we observe a force acting on a superconducting drum resonator integrated in a microwave optomechanical cavity through the nonlinear dynamics this force imparts to the resonator. The measured dynamics points to an extremely intense force found to be compatible in magnitude with the Casimir force for the range of vacuum separations that can be expected in this device, and incompatible with estimates of other known sources of nonlinearity. 
This nonlinearity is intense enough that, with a modified design, this device type should operate in the single-phonon nonlinear regime. Accessing this regime has been a long-standing goal that would greatly facilitate quantum operations of mechanical resonators.
\end{abstract}

\date{\today}
\maketitle

\section*{Introduction}
A significant body of theoretical and experimental works over the last 20 years has focused on understanding the Casimir force~\cite{Casimir1948,Lifshitz1956,Lamoreaux1997,Mohideen1998,Rodriguez2011,Woods2016,Gong2021}.
This is motivated in part by the Casimir effect's practical applications in MEMS~\cite{Serry1998,Chan2001,Broer2013,Tang2017,Stange2019,Pate2020}, but to a larger extent by its intimate connection to heat and energy transfer~\cite{Fong2019,Xu2022}, especially at finite conductivity and temperature~\cite{Bezerra2004,Hoye2007,Sushkov2011,Mostepanenko2021}. Furthermore, Casimir force experiments have also put bounds on Yukawa corrections to short-distance Newtonian gravity~\cite{Decca2003,Klimchitskaya2012} and the Casimir effect has been proposed as a candidate to account for dark energy in cosmology~\cite{Leonhardt2019}. Recently, the interplay between the Casimir force and superconductivity has garnered interest, either manifesting as a correction to the superconducting condensation energy~\cite{Bimonte2005,PerezMorelo2020}, or as a correction to the electric permittivity of the material~\cite{Bimonte2010,Norte2018,Rodrigues2018,Bimonte2019}. Studies of this interplay could shed light on the discrepancy between the calculated and experimentally observed Casimir force between real metals at finite temperature~\cite{Decca2007,Bimonte2019,Mostepanenko2021}. While most room-temperature experiments rely on a variation of the distance between objects to observe the expected force-distance scaling~\cite{Lamoreaux1997,Mohideen1998,Andrews2015,Tang2017,Stange2019}, it is challenging to perform this type of experiment with superconductors due to the difficulty of precise positioning in cryogenic environments. By contrast, the strong nonlinearity imparted by the Casimir potential can be easily shown through the dynamical behavior of an oscillator without relying on precise positioners. Only few works have studied the Casimir force through the nonlinear dynamics~\cite{Serry1995,Chan2001a}. 

Strong nonlinearities in oscillators at cryogenic temperatures are especially interesting in the context of resonators in the quantum regime. Cavity optomechanics, where the motion of an object couples to the electromagnetic field of a cavity, has allowed to engineer Gaussian quantum mechanical states in cryogenic mechanical oscillators. Notably, resonators were cooled to the ground state~\cite{Teufel2011}, their uncertainty was squeezed below the magnitude of quantum fluctuations~\cite{Wollmann2015,Pirkkalainen2015,Lecocq2015}, entangled states of two oscillators were prepared~\cite{OckeloenKorppi2018,Kotler2021}, and mechanical states were measured avoiding quantum backaction~\cite{Mercierdelepinay2021}. However, linear optomechanical systems cannot realize non-Gaussian quantum states such as Fock states, cat states, and entangled combinations thereof, which are desirable for quantum algorithms. To realize these states nonetheless, mechanical oscillators have been addressed with non-Gaussian photonic states~\cite{Riedinger2016,Galinskiy2024}, or coupled to superconducting qubits~\cite{OConnell2010,Chu2018,Marti2024,Yang2024} or quantum dots~\cite{Samanta2023} to inherit their nonlinearity. In contrast, the nonlinearity imparted by the Casimir force to a superconducting drum resonator is intrinsic to the construction of the system and does not rely on coupling to another system.

In this work, we have observed a nonlinear potential that we show to be quantitatively compatible with the Casimir potential acting between the plates of a type-I superconducting drum. This force force yields a strong softening nonlinearity, where the resonating frequency decreases with increasing amplitude~\cite{Serry1995,Chan2001a,Chan2001}. The optomechanical coupling to a microwave cavity  allows us to calibrate all properties of the system and compare its response to that obtained within a model of Casimir potential with only a single fit parameter. We qualitatively and quantitatively exclude other (nonlinear) effects that have hindered earlier works~\cite{Andrews2015,Eerkens2017,Norte2018}. This system requires no cryogenic positioners, and it shows a strong, tunable mechanical nonlinearity that is fully compatible with a mechanical system where quantum operations have been achieved. 

\begin{table}
\renewcommand{\tablename}{\textbf{Table}}
\begin{tabular}{l|l|l}
Quantity & Symbol & Value \\
\hline
Unperturbed mech.~frequency** & $\omega_\mathrm{r}$ & $2\pi \times 16.247$~\si{\mega\hertz} \\
Mechanical frequency* & $\omega_\mathrm{m}$ & $2\pi \times 10.001$~\si{\mega\hertz} \\
Mechanical linewidth* & $\gamma_\mathrm{r} \!\simeq\!\gamma_\mathrm{m}$ & $2\pi \times 168.9 \pm 9.5
$~\si{\hertz} \\
Cavity frequency* & $\omega_\mathrm{c}$ & $2\pi \times 5.461831$~\si{\giga\hertz} \\
External linewidth* & $\kappa_\mathrm{e}$ & $2\pi \times 250.8 \pm 2.1
$~\si{\kilo\hertz} \\
Internal linewidth* & $\kappa_\mathrm{i}$ & $2\pi \times 297.2 \pm 2.8
$~\si{\kilo\hertz} \\
Drive frequency & $\omega_\mathrm{mw}$ & $\simeq \omega_\mathrm{c}$ \\
Sideband frequency* & $\omega_\mathrm{sb}$ & $\simeq \omega_\mathrm{mw} - \omega_\mathrm{m}$ \\
Detuning $\omega_\mathrm{c} - \omega_\mathrm{mw}$ & $\Delta$ & $< |2\pi \times 1|$~\si{\kilo\hertz} \\
Optomechanical coupling* & $g_0$ & $2\pi \times 150 \pm 9
$~\si{\hertz} \\
Effective mass & $m_\mathrm{eff}$ & $3.96 \times 10^{-14}$~\si{\kilo\gram} \\
Vacuum gap w/o Cas. force** & $d$ & $18.00 \pm 0.25$~\si{\nano\meter} \\
Vacuum gap** & $d'$ & $15.10 \pm 0.25$~\si{\nano\meter} \\
Bottom plate diameter & $2r$ & $11.3$~\si{\micro\meter} \\
Moving plate diameter & - & $14.5$~\si{\micro\meter} \\
Casimir pressure at rest** & $P_\mathrm{c}$ & $12.0 \pm 0.6$~\si{\kilo\pascal} \\
Casimir force amplitude** & $P$ & $(1275 \pm 7) \cdot 10^{-24}~\mathrm{Pa}~\mathrm{m}^{n}$\\
Cas.~force scaling exponent** & $n$ & $3.193$ \\

\end{tabular}
\caption{\textbf{The parameters of the system} either from design, calibrated at small mechanical amplitudes using optomechanical models (*), or estimated within the Casimir force model (**). Intervals represent $95\%$ confidence.}
\label{Tableparameters}
\end{table}

\section*{Results}
\subsection*{The nonlinear dynamics}
\begin{figure*}[ht]
\renewcommand{\figurename}{\textbf{Fig.}}
\includegraphics[width = \textwidth]{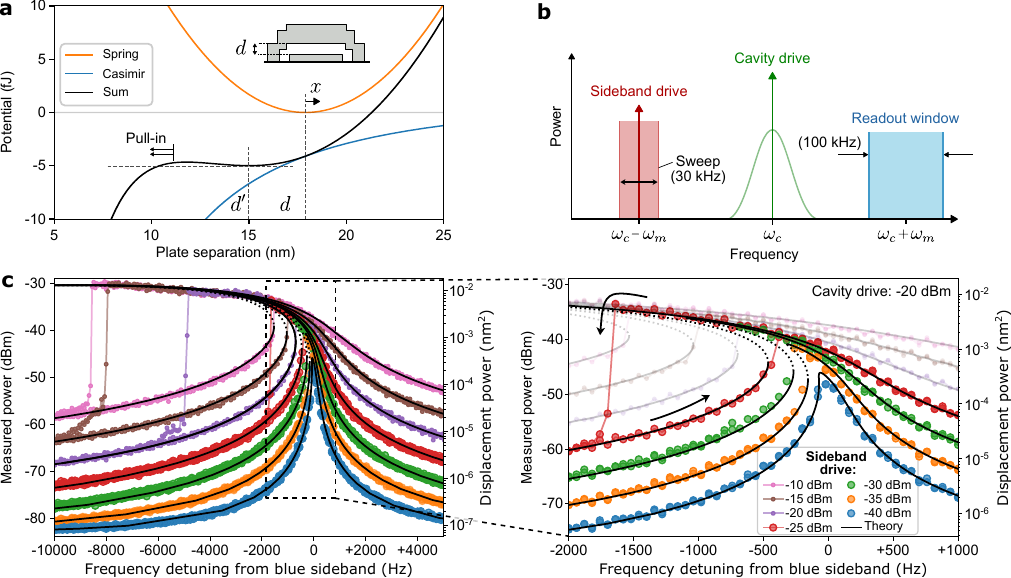}
\caption{\textbf{Effects of a strong nonlinear potential on a superconducting drum resonator. a}: Our device consists of two plates separated by a vacuum gap. 
The top plate is mechanically compliant and experiences a mechanical restoring potential (orange line) which is minimum for a separation $d$ between plates (see inset). The nonlinear potential (blue) adds to this restoring potential, and the sum (black) has a local minimum at separation $d'$ shifted from $d$. The total potential is not harmonic around $d'$ for large amplitudes which causes nonlinear behavior. \textbf{b}: Schematic of the microwave optomechanical measurement scheme. We strongly drive the microwave cavity at its center frequency, with an additional strong drive tone at $\omega_\mathrm{sb}$ which is close to $\omega_\mathrm{c} - \omega_\mathrm{m}$. The sideband drive is swept in frequency, and we read out the emitted signal in a window around $\omega_\mathrm{c} + \omega_\mathrm{m}$. \textbf{c}:  
Measured displacement response for various nominal sideband drive powers (labels) at $-20$~dBm cavity drive, showing the expected Lorentzian behavior at low drive power that transitions to a strong softening nonlinearity at high drive power. The Casimir force model (black line) becomes multi-valued, with two stable branches (solid black) and an unstable solution (dotted black/semitransparent). Our measurements follow the stable solutions, leading to a hysteresis depending on the sweep direction. The theory matches well to all curves using only a single fit parameter, $d$, common to all curves. The right panel highlights details of the center region indicated by the dashed box, and the hysteresis depending on the sweep direction is indicated by arrows. The data points at the highest powers are made transparent for visual clarity.}
\label{FigSchematicCasimir}
\end{figure*}

Here, we describe the experimental system. The drum resonator consists of two plates made of evaporated aluminium. The top plate is suspended and forms a harmonic oscillator of frequency $\omega_r$: the minimum of the harmonic mechanical potential would correspond to a vacuum gap of $d$. However, observations detailed below indicate that the oscillator is subject to a very strong nonlinear attractive potential. Such a potential is expected to draw the mechanically compliant top plate closer to the bottom plate that is fixed to a rigid substrate, modifying the vacuum gap to a new value $d'$, which we schematically show in Fig. \ref{FigSchematicCasimir}a. Furthermore, the frequency of oscillations around this local potential minimum is expected to be lowered by the nonlinear potential: we denote $\omega_m$ the value of the softened frequency. Modified gap $d'$ and frequency $\omega_m$ are the parameters physically realized by the device since the nonlinear potential exists throughout all experiments, and parameters $d$ and $\omega_r$, that concern a hypothetical situation where the drum would not be subject to a nonlinear potential, are only introduced as convenient modelling intermediates. The oscillator is expected to stably oscillate at $\omega_m$ around its local potential minimum with a small amplitude, but at larger amplitudes the frequency is expected to decrease (spring softening) until it reaches the pull-in point, beyond which the top plate would collapse onto the bottom plate~\cite{Serry1998,Rodrigues2018}.

The drum resonator is mounted 
in a dilution cryostat and  stabilized at \SI{10}{\milli\kelvin} (see \ref{Methods}), so the aluminium is deep in the superconducting regime. The drum forms the variable capacitance of a microwave cavity, which we drive with two strong drives, one at the resonance frequency of the microwave cavity, $\omega_\mathrm{c}$ (cavity drive), and the other at the red sideband $\omega_\mathrm{c} - \omega_\mathrm{m}$ (sideband drive) as shown schematically in Fig.~\ref{FigSchematicCasimir}\textbf{b}. This combination of tones generates a driving force on the mechanical resonator such that the mechanical amplitude scales linearly with either drive power (see Supplementary Sec.~\ref{SISecOptomechanics}). All drive powers are reported as their nominal values at room temperature, they are divided by attenuators before reaching the sample as shown in \ref{Methods}. The cavity drive power is much larger than the sideband drive, and we sweep the frequency of the weaker sideband drive within a range of \SI{30}{\kilo\hertz} both upwards and downwards in frequency. We record the  output signal of the cavity with a spectrum analyzer over a window centered around $\omega_\mathrm{c} + \omega_\mathrm{m}$ (blue sideband). 

The signal we observe at the readout sideband is proportional to the displacement of the resonator. At low drive powers, we see a Lorentzian response compatible with a mechanical resonance frequency $\omega_m\simeq 10.0\,\rm MHz$, but increasing the cavity drive power in Fig.~\ref{FigSchematicCasimir}\textbf{c} shows a strong softening nonlinearity. At high drive powers, this nonlinearity creates two stable solutions to Eq.~\eqref{EqCasimirEOM} over a limited range of frequencies (a low-amplitude and a high-amplitude branch) with an unstable solution in between. The sweep direction determines which branch is followed as indicated by arrows in Fig.~\ref{FigSchematicCasimir}\textbf{c}. The resonator amplitude jumps up at the end of the low-amplitude branch (bifurcation point), but it jumps down from the high-amplitude branch at a point before the end of the branch. This occurs due to our stepped sweep protocol (see \ref{Methods}).

\subsection*{A Casimir force model}
A reasonable candidate to explain the strong nonlinearity  is the Casimir effect~\cite{Chan2001}, which is expected to be intense in these devices given their small vacuum gap values ($15-50\,\rm nm$), relatively large surfaces and high susceptibilities. As explained further below, a model for the dynamics of the oscillator in presence of the Casimir force is 
\begin{equation}
\ddot{x} + \gamma_\mathrm{r} \dot{x} + \omega_\mathrm{r}^2 x + \frac{P}{(x + d)^n } \frac{\pi r^2 }{m_\mathrm{eff}} = \frac{F_0}{m_\mathrm{eff}},
\label{EqCasimirEOM}
\end{equation}
which describes the evolution of a point-like harmonic oscillator in one dimension with displacement coordinate $x$, resonance frequency $\omega_\mathrm{r}$, effective mass $m_\mathrm{eff}$, and decay rate $\gamma_\mathrm{r}$, optomechanically driven by a force $F_0$. It is additionally subject to a force derived from the Casimir pressure, the last term on the left-hand side, that works on the area of the drum, $\pi r^2$. This model reducing the distributed motion of the membrane to a scalar coordinate with an effective mass stays roughly valid in presence of a nonlinear potential. We compute the Casimir pressure for the real material based on the Mattis-Bardeen conductivity for (BCS) superconductors
\cite{Bimonte2019}, see \ref{Methods}. From these computations, the Casimir force would have an expression of the form $P_\mathrm{c} = \frac{P}{(x + d)^{n}}$ where $P \simeq 1275 \cdot 10^{-24} ~\mathrm{Pa}\cdot\mathrm{m}^n$ and pressure-distance scaling $n=3.193$. This power law is the one predicted by the Lifshitz formalism in the range of separation gaps of the device. The use of this power law here, as well as the reference to the Casimir force rather than the van der Waals force, is motivated by the fact that the order of magnitude of the plates separation is larger than the threshold separating the Casimir regime from the van der Waals regime where retardation effects are negligible (see Supplementary Sec.~\ref{SISecCasimirvdW} for more details).

\begin{figure}
\renewcommand{\figurename}{\textbf{Fig.}}
\includegraphics[width = 0.5\textwidth]{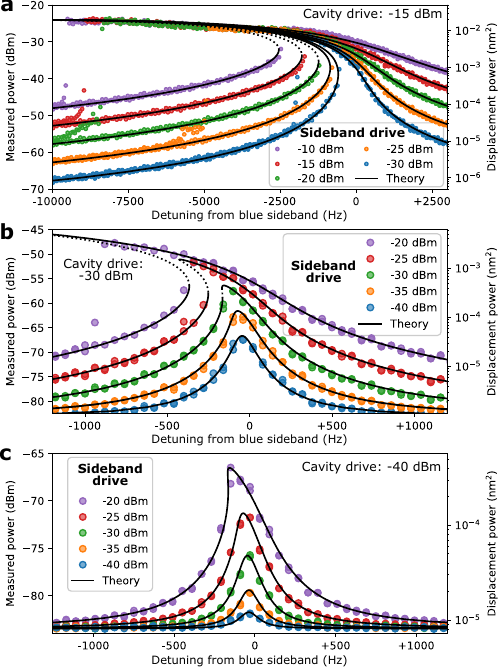}
\caption{
\textbf{Responses measured for all combinations of drive parameters}
for cavity drive powers (\textbf{a}: $-15$~dBm, \textbf{b}: $-30$~dBm, \textbf{c}: $-40$~dBm) and sideband drive powers (colors). 
All theory curves (black lines) across all panels share only a single fit parameter $d$.}
\label{FigAllCasimircurves}
\end{figure}

We compute the expected displacement of the oscillator of Eq.~\eqref{EqCasimirEOM} for a frequency-swept drive (see ~\ref{Methods} and \ref{SI}) using the parameters of Table ~\ref{Tableparameters}, and adjust a single parameter, $d$, to collectively fit all measured mechanical responses. The resulting curves are plotted as black lines in Fig.~\ref{FigSchematicCasimir}\textbf{c}, with the unstable part of the solution represented as a dotted line. This procedure yields  an estimate of the hypothetical unperturbed gap $d=18\,\rm nm$, giving an excellent agreement for all measured data simultaneously.
We repeat the experiment and analysis at different combinations of cavity and sideband drive powers, shown in Fig.~\ref{FigAllCasimircurves}: the same model and fit parameter accurately describe the shape of the signal across more than three orders of magnitude in displacement, up to a maximum measured amplitude of $\simeq\!100$~\si{\pico\meter}.

Within this modelling, the shifted equilibrium position of the oscillator is $d' = 15.10 \pm 0.25$~\si{\nano\meter}, corresponding to a static pull $x =d'-d \simeq -2.9\,\rm nm$ under the nonlinear potential. The inferred physical gap $d'$ is of the expected order of magnitude for this device. The value of the unperturbed gap $d$ is furthermore in very good agreement with independent simulations of the shape of the membrane cooled to cryogenic temperature in absence of a nonlinear potential (see ~\ref{Methods}). The unperturbed mechanical frequency $\omega_r$ is furthermore found to be $\omega_r\simeq 16.2\,\rm  MHz$ within this model, which is also in good agreement with models of the membrane, indicating a nonlinear softening larger than $6\,\rm MHz$ to obtain the physical frequency $\omega_m\simeq 10.0\,\rm MHz$. 
The total pressure exerted on the membrane at rest is estimated to be as high as $P_c \simeq 12.0 \pm 0.6$~\si{\kilo\pascal}, which leads to the considerable static displacement and frequency pull modelled.

\subsection*{Calibration of displacement amplitude}
\begin{figure}[ht]
\renewcommand{\figurename}{\textbf{Fig.}}
\includegraphics[width = 0.5\textwidth]{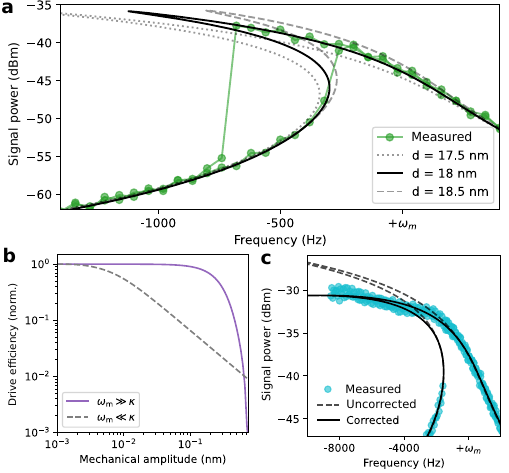}
\caption{\textbf{Optomechanical calibration. a}: Theory curves generated with equal maximum mechanical amplitude for various values of $d$ (black \& grey lines). For smaller $d$, the Casimir force is stronger and leads to a larger softening nonlinearity. For $d = 18$~\si{\nano\meter} there is excellent agreement with the measured data (green). \textbf{b}: The drive efficiency decreases sharply from 1 at high amplitude, as most of the time the instantaneous cavity frequency $\omega_\mathrm{c}(t)$ is far away from the drive frequency. This effect is stronger, comparatively, in the bad cavity limit ($\omega_\mathrm{m} \ll \kappa$), but it is noticeable in our experiments at the largest amplitudes. \textbf{c}: The drive efficiency leads to a correction of the simulated response curves at large amplitudes.}
\label{FigOptomechanicsCasimir1}
\end{figure}

To make $d$ (the vacuum gap at the minimum of the harmonic mechanical potential) the only fit parameter, we calibrate all other parameters listed in Table \ref{Tableparameters} to accurately infer the displacement amplitude of the resonator. The drum resonator is embedded in an optomechanical cavity such that its position couples to the radiation pressure. 
The equation of  evolution for the cavity field is
~\cite{Aspelmeyer2014}
\begin{equation}
\dot{{a}}(t)    = -(i\Delta + \kappa/2){a} + ig_0 \tilde{x}{a} + \sqrt{\kappa_\mathrm{e}}{s}_\mathrm{in}(t).
\label{EqOptomechanicalEOM}
\end{equation}
Here, $\tilde{x}$ represents the displacement around the equilibrium position shifted by the nonlinear potential and ${a}$ is the cavity field amplitude. We work in a frame rotating at the frequency of the cavity drive, which is detuned by $\Delta$ from the cavity resonance frequency. The cavity linewidth $\kappa = \kappa_\mathrm{e} + \kappa_\mathrm{i}$ is the sum of the external and internal decay rates (see Table \ref{Tableparameters}). The optomechanical single-photon coupling is $g_0$. 
The input coherent signal is modelled by ${s}_\mathrm{in}$. The mechanical and cavity noise are negligibly small compared to driven amplitudes in this problem and we exclude them from the description. 

The equation of motion, Eq.~\eqref{EqOptomechanicalEOM}, depends on parameters $g_0$, $\kappa$, and $\Delta$. To calibrate the measured signal into a displacement, we complete this equation with the small-displacement motion equations around the equilibrium position (see \ref{SI}), which additionally depend on $\omega_m$ and $\gamma_m$ the mechanical parameters in the local minimum of potential.
We first extract cavity parameters $\kappa_\mathrm{e}$ and $\kappa_\mathrm{i}$ from the cavity reflection response, and get the single-photon coupling $g_0$, the mechanical parameters $\omega_m, \gamma_m$ and the power at our detector from a given displacement power from the measurement of thermal motion of the drum resonator at \SI{10}{\milli\kelvin} (see separate \ref{SI} section on these preliminary calibrations). Then, using these independently calibrated values, we use a numerical solution of the equations of motion to calibrate the ratios between the powers of the tones at the output of generators and incoming on the device. These powers enter into ${s}_{\rm in}(t)$ in Eq.~\eqref{EqOptomechanicalEOM}: they determine the driving force $F_0$ in model \eqref{EqCasimirEOM} and thereby the driven mechanical amplitude. The numerical solution for the cavity signal at the first two sidebands (red and blue) is matched to the measured signal by adjusting the power ratios, completing the calibration.

Parameters $P$ and $n$ are related to material parameters for aluminium for which we can obtain a high accuracy (see \ref{Methods}). This makes $d$ the sole fit parameter to compare the solution of Eq.~\eqref{EqOptomechanicalEOM} to our experimental observations. In Fig.~\ref{FigOptomechanicsCasimir1}\textbf{a}, we show that $d = 18$~\si{\nano\meter} generates the best-fitting solution. Also solutions with $d = 17.5,~18,$ and $18.5$~\si{\nano\meter} are overlaid on experimental data obtained at $-20$~dBm cavity drive and $-30$~dBm sideband drive. The fit procedure is applied to all measured response signals (see Fig.~\ref{FigAllCasimircurves}). From this collective fit, we estimate the uncertainty of $d$ as the upper bound of the uncertainty propagated from the amplitude calibration and the standard error of the fitted response signal. The uncertainty we find this way is larger than the uncertainty in $d$ we would have found from variations in parameters $P$ and $n$, which motivates our choice to treat them as fixed values.

\subsection*{Drive efficiency}
At the core of Eq.~\eqref{EqOptomechanicalEOM} is the notion that the cavity frequency depends on the mechanical position, $\omega_\mathrm{c}(x)$. For a large enough displacement amplitude $x$, the cavity frequency shifts by more than the cavity linewidth $\kappa$ such that the cavity drive tone is mostly reflected. This reduces the amount of circulating power in the cavity, and it reduces the effectiveness of the drive on the mechanical resonator. We estimate this loss of efficiency and compensate for it in the following way. We define the `drive efficiency' as the circulating power in the cavity at some mechanical amplitude normalized to the power at low displacement. It is straightforward to calculate by solving Eq.~\eqref{EqOptomechanicalEOM} for fixed amplitude of mechanical motion, and the results are plotted in Fig.~\ref{FigOptomechanicsCasimir1}\textbf{b}. In the experiment $\omega_\mathrm{m}\gg \kappa$ so the drive efficiency is close to 1 up to $\sim 100$~\si{\pico\meter} and strongly decreases beyond that.

The effect of the drive efficiency is shown in Fig.~\ref{FigOptomechanicsCasimir1}\textbf{c}. For the frequencies where the mechanical amplitude is small, the uncorrected and corrected curves overlap since the drive efficiency is 1. At large amplitudes the uncorrected prediction overestimates the measured response ($-20$~dBm cavity drive and $-10$~dBm sideband drive).

\subsection*{Higher-order sidebands scattering}
\begin{figure}[ht]
\renewcommand{\figurename}{\textbf{Fig.}}
\includegraphics[width = 0.5\textwidth]{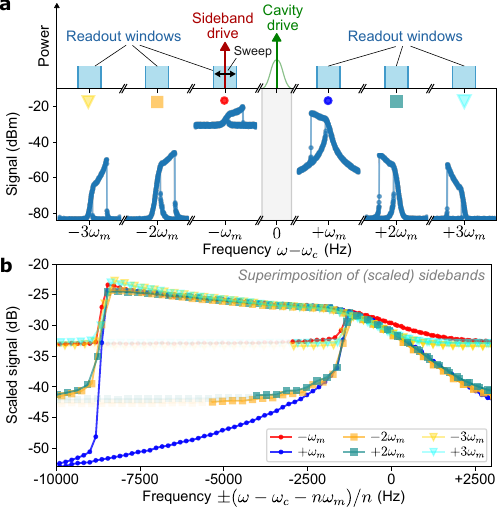}
\caption{\textbf{Response observed at higher-order sidebands. a}: The full optomechanical measurement consists of six sidebands that are read out sequentially. The first red sideband ($-\omega_\mathrm{m}$) is at the same frequency as our swept sideband drive, so its signal is superposed on a pedestal from the directly reflected sideband drive signal. \textbf{b}: Superimposition of sidebands. All sidebands encode 
the mechanical displacement
with a known proportionality (see main text). The frequency axis of each sideband has been shifted to align the sideband frequencies $\omega_c+n\omega_m$, and the frequency axis of the red sidebands (below the cavity frequency) has been flipped. Furthermore the frequency range has been divided by two for second sidebands and three for third sidebands.
The vertical axis of $2^\mathrm{nd}$ and $3^\mathrm{rd}$ order sidebands has been squared and cubed respectively, as expected from theory.
}
\label{FigOptomechanicsCasimir2}
\end{figure}

The optomechanical interaction, Eq.~\eqref{EqOptomechanicalEOM}, is intrinsically nonlinear and it is here driven by large cavity fields and large mechanical amplitudes. As a result, scattering between sidebands assisted by the mechanical displacement yields multiple equally-spaced peaks in the observed response. 
We read out six of these sidebands, at $\omega_\mathrm{c} \pm n \times \omega_\mathrm{m}$ ($n = 1,~2,~3$), as shown in Fig.~\ref{FigOptomechanicsCasimir2}\textbf{a}. The higher-order sidebands can be viewed as repeated scattering events. We extend classical scattered mode theory~\cite{Kippenberg2007} to cover all six sidebands. We write the system of linear coupled equations for the classical field amplitudes $a_n$ at the corresponding frequency components in matrix form,
\begin{equation}
\begin{bmatrix}
1 & d_{+3} & 0 & 0 & 0 & 0 & 0 \\
d_{+2} & 1 & d_{+2} & 0 & 0 & 0 & 0 \\
0 & d_{+1} & 1 & d_{+1} & 0 & 0 & 0 \\
0 & 0 &d_{0} & 1 & d_{0} & 0 & 0 \\
0 & 0 & 0 & d_{-1} & 1 & d_{-1} & 0 \\
0 & 0 & 0 & 0 & d_{-2} & 1 & d_{-2} \\
0 & 0 & 0 & 0 & 0 & d_{-3} & 1
\end{bmatrix}
\begin{bmatrix}
a_{+3} \\
a_{+2} \\
a_{+1} \\
a_{0} \\
a_{-1} \\
a_{-2} \\
a_{-3}
\end{bmatrix}
= 
\begin{bmatrix}
0 \\
0 \\
0 \\
\frac{a_\mathrm{mw}\sqrt{\kappa_\mathrm{e}}}{-i\Delta + \kappa/2} \\
0 \\
0 \\
0
\end{bmatrix}
\label{EqSidebandmatrix}
\end{equation}
with
\begin{equation}
d_{\pm n} = \frac{g_0 x_{\rm amp}}{2 x_\mathrm{zpf}} \frac{1}{-i(\Delta \pm n\omega_\mathrm{m}) + \kappa/2}.
\end{equation}
Here $a_\mathrm{mw}$ is the input power of the microwave drive at the cavity frequency, $a_\mathrm{sb}$ is the input power of the microwave drive at the red sideband. We cut off the scattering processes above order $|n|=3$, retaining 7 frequency ranges in Eq.~\eqref{EqSidebandmatrix}. This is a valid approximation when $g_0 \ll \kappa/2$, since each individual photon is much more likely to exit the cavity than to scatter from the mechanical resonator. For more details, see the \ref{SI}. The power in the first-order sidebands ($\pm \omega_\mathrm{m}$) scales linearly with the amplitude of mechanical displacement ${x}_{\rm amp}$, while the second order sidebands ($\pm 2\omega_\mathrm{m}$) scale quadratically and the third order sidebands ($\pm 3 \omega_\mathrm{m}$) scale cubically. The span of the sweep is doubled and tripled for the second and third mode. When we combine these known behaviors and flip the frequency of the $-1,-2,-3\times\omega_\mathrm{m}$ sidebands, we can 
overlay all sidebands in the high-amplitude regime (Fig.~\ref{FigOptomechanicsCasimir2}\textbf{b}).

\section*{Discussion}
There are many other effects that could result in an apparent nonlinearity, such as the electrostatic force between the drum plates, potential patches in the aluminium, the geometric nonlinearity, the nonlinear optomechanical coupling, and superconducting vortices. We can rule out these effects as an alternative explanation for our observations, as described in detail in the \ref{SI}. Put briefly, we estimate the pressure contributed by e.g., potential patches, based on Kelvin probe force measurements of the electrostatic potential. We simulate the effect of patches in our geometry~\cite{deJong2024}, and find that the patch pressure is several orders of magnitude smaller than the Casimir pressure at our plate separation. Based on this, we rule out the effect of potential patches (and similar for the other effects described in the \ref{SI}). 
We have not found another source on nonlinearity that could mimic the effect of the Casimir force in this device and consider it to be a strong candidate to explain the observations.

Our future experiments will involve a gate electrode patterned around the drum plate to allow electrostatic tuning of the separation~\cite{Andrews2015} and, by extension, the Casimir force or nonlinear potential. A modest \SI{100}{\pascal} electrostatic pressure is achievable, and sufficient to merge the locally stable solution used in this work with the pull-in point. This should allow for nonlinear mechanics on the single-phonon level, a long-standing goal previously only achieved by coupling to external systems~\cite{Samanta2023,Yang2024}. The gate electrode will also allow us to improve the accuracy of the amplitude calibration to improve our bounds on $d$ and $P_\mathrm{c}$, to shed light on the Casimir puzzle~\cite{Mostepanenko2021} of materials with finite conductivity.

In summary, we have observed strongly nonlinear dynamics of a superconducting membrane compatible with the existence of a Casimir potential between superconducting plates. We have shown that this effect does not seem to correspond to other known sources of nonlinearity.  
Due to the exceptionally small spacing between plates, the Casimir force in this configuration is expected to be much larger than in conventional Casimir experiments, including those at room temperature: for comparison, it is expected to be of the order of magnitude of a tenth of the atmospheric pressure. This suggests that our system is well-suited to accurately probe the remaining uncertainties around the Casimir effect. By adding a gate electrode, we could improve the precision of our nonlinear-dynamics-based measurements beyond the small predicted change at the superconducting transition~\cite{Bimonte2019}. More fundamentally, 
our experimental setup gives a unique access to an extremely strong, tuneable nonlinearity that is intrinsic to the mechanical system. This means that we could obtain well-separated mechanical energy levels without resorting to the nonlinearity of another system, i.e., bring the mechanical resonator into quantum regime without interactions with non-Gaussian photon states or coupling to an external nonlinear system. Furthermore, the competition between the nonlinear potential and the elastic potential can be tuned electrostatically, in situ, to be locally flat, which could be used to probe macroscopic mechanical quantum tunneling~\cite{Sillanpaa2011} or make a mechanical qubit~\cite{Pistolesi2021,Yang2024}.\\

\section*{Methods}\customlabel{Methods}{Methods}
\subsection*{Calculation of the Casimir force}
\label{SecSICasimirforcedistance}
To compute the Casimir force between the superconducting plates of our drum, we follow exactly the method of Ref~\cite{Bimonte2019}. It is based on the Lifshitz formula~\cite{Lifshitz1956} for the pressure $P(d,T)$ between two plates as a function of their separation $d$ and temperature $T$,
\begin{widetext}
\begin{equation}
P(d,T) = \frac{-k_B T}{\pi}\sum_0^\infty\int_0^\infty dk_\perp k_\perp q_\ell
\left( \left[ \frac{e^{2dq_\ell}}{r_\mathrm{TE,1}(i\xi_\ell,k_\perp)r_\mathrm{TE,2}(i\xi_\ell,k_\perp)} -1 \right] ^{-1} + 
\left[ \frac{e^{2dq_\ell}}{r_\mathrm{TM,1}(i\xi_\ell,k_\perp)r_\mathrm{TM,2}(i\xi_\ell,k_\perp)} -1 \right] ^{-1} \right),
\label{EqCasimirPressure}
\end{equation}
\end{widetext}
where $k_\perp$ is the in-plane momentum, the $\ell = 0$ term in the sum has a weight of one half, $\xi_\ell = 2\pi \ell k_B T/\hbar$ are the imaginary Matsubara frequencies, $q_\ell = \sqrt{\xi_\ell^2/c^2 + k_\perp^2}$, and TE and TM indicate the two independent polarizations of the electromagnetic field. The Fresnel reflection coefficients for the polarizations (TE,TM) and plates (1,2) are~\cite{Bimonte2019}
\begin{equation}
\begin{aligned}
r_\mathrm{TE}(i\xi_\ell,k_\perp) &= \frac{q_\ell - s_\ell}{q_\ell + s_\ell}, \\
r_\mathrm{TM}(i\xi_\ell,k_\perp) &= \frac{\epsilon_\ell q_\ell - s_\ell}{\epsilon_\ell q_\ell + s_\ell},
\end{aligned}
\end{equation}
where $s_\ell = \sqrt{\epsilon_\ell \xi_\ell^2/c^2 +k_\perp^2}$ and electric permittivity $\epsilon_\ell = \epsilon(i\xi_\ell)$ for the materials of each of the plates 1 and 2. For normal-state materials, we use the Drude model dielectric function $\epsilon_\mathrm{Drude}$. For superconductors, the Mattis-Bardeen fomula gives a corrected function $\epsilon_\mathrm{BCS}$. The analytic continuation of both functions has been derived as~\cite{Bimonte2010,Bimonte2019}
\begin{equation}
\begin{aligned}
\epsilon_\mathrm{Drude}(i\xi) &= \epsilon_0 + \frac{\Omega_{p}^2}{\xi(\xi + \gamma_\mathrm{p})}, \\
\epsilon_\mathrm{BCS}(i\xi) &= \epsilon_0 + \frac{\Omega_{p}^2}{\xi}\left(\frac{1}{\xi + \gamma_\mathrm{p}} + \frac{g(\xi,T)}{\xi}\right).
\end{aligned}
\end{equation}
Here, $\Omega_\mathrm{p}$ represents the plasma frequency for intraband transitions, $\gamma_\mathrm{p}$ is the relaxation frequency. The contribution from BCS theory is in the form of the factor $g(\xi,T)$, given as~\cite{Bimonte2019}
\begin{equation}
g(\xi,T) = \int_{-\infty}^\infty \frac{d\epsilon}{E} \tanh\left( \frac{E}{2k_B T}\right) \mathrm{Re}\left[G_+(i\xi,\epsilon)\right].
\end{equation}
This expression is valid for temperatures below the superconducting transition temperature, $T<T_\mathrm{c}$. The function $G_+$ is defined as
\begin{equation}
\begin{aligned}
G_+(z,\epsilon) &= \frac{\epsilon^2 Q_+(z,E) + \left[Q_+(z,E)+i\hbar\gamma\right]A_+(z,E)}{Q_+(z,E)\left( \epsilon^2 - \left[Q_+(z,E) + i\hbar\gamma \right]\right)}, \\
E &= \sqrt{\epsilon^2 + \Delta^2}, \\
Q_+(z,E) &= \sqrt{(E + \hbar z)^2 - \Delta^2}, \\
A_+(z,E) &= E(E + \hbar z) + \Delta^2,
\end{aligned}
\end{equation}
and the superconducting gap $\Delta(T)$ is temperature dependent,
\begin{equation}
\Delta = c_1 k_B T_\mathrm{c} \sqrt{1 - \frac{T}{T_\mathrm{c}}}\left(c_2 + c_3 \frac{T}{T_\mathrm{c}}\right),
\end{equation}
where we take $c_1 = 1.764$, $c_2 = 0.9963$, and $c_3 = 0.7735$ from BCS theory~\cite{Tinkham1996}.

To evaluate the expression of Eq.~\eqref{EqCasimirPressure}, we need the material parameters for aluminium, which are tabulated~\cite{Palik1997}. We use $\Omega_\mathrm{p} = 13~\mathrm{eV/\hbar}$, $\gamma_\mathrm{p} = \gamma_0/\mathrm{RRR}$ where $\mathrm{RRR} = 2$ denotes the residual resistance ratio of our thin-film aluminium, and $\gamma_0 = 0.1~\mathrm{eV/\hbar}$ is a phenomenological relaxation rate (dissipation of current). Finally, setting $\epsilon_0 =1.03$ as the relative permittivity of aluminium takes into account the core interband transitions~\cite{Palik1997}.

To numerically compute the Casimir pressure in a reasonable amount of time, we take the Matsubara frequencies up to $\ell = 200,000$. We determine that this is sufficient by increasing the range of the sum until the final pressure changes by less than $10^{-5}$ of the value for the previous range. Conversely, to compute the difference between $\epsilon_\mathrm{Drude}$ and $\epsilon_\mathrm{BCS}$ it is sufficient to take~\cite{Bimonte2019} $\ell \lesssim 600$. Nonetheless, we must also integrate over $k_\perp$. The integrand of Eq.~\eqref{EqCasimirPressure} is sharply peaked~\cite{Bimonte2010}, and heuristic bounds of $\pm 300\sqrt{\xi_\ell \Delta}$ seem to have sufficiently small error. 

\begin{figure}
\renewcommand{\figurename}{\textbf{Fig.}}
\includegraphics[width = 0.5\textwidth]{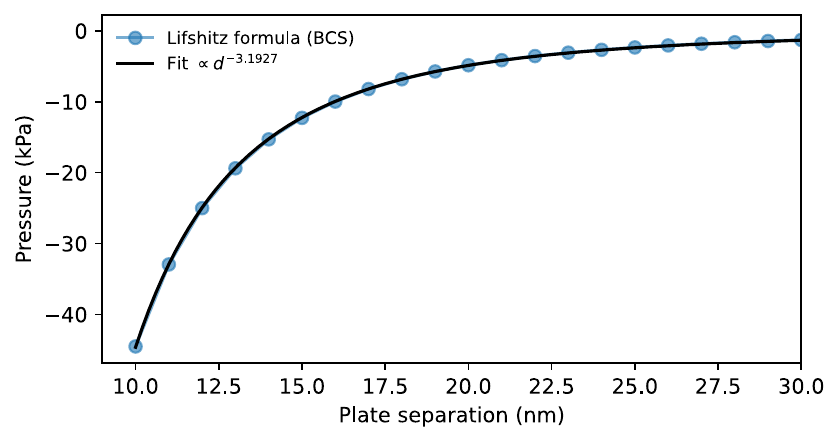}
\caption{\textbf{The Casimir pressure.} The Casimir pressure for various distances as calculated via the Lifshitz formula and the BCS model, and a fit proportional to $d^{-3.193}$.}
\label{FigCasimirPressure}
\end{figure}

In Fig.~\ref{FigCasimirPressure}, we plot the results of our evaluation of Eq.~\eqref{EqCasimirPressure} for our system. While the Casimir pressure scales as $P\propto d^{-4}$ for ideal conductors~\cite{Casimir1948}, for real conductors the scaling exponent is different. Using the material parameters for aluminium, the pressure is best described by a fit $P_\mathrm{c} = - \frac{1275\pm 7 \cdot 10^{-24}}{d^{3.193}}$~\si{\pascal}, with the $95\%$ confidence intervals extracted from the fit to the calculation results. We have plotted the dependence $P\propto d^{-3.193}$ as a black line in Fig.~\ref{FigCasimirPressure}.

\subsection*{Drum design and fabrication}\label{SecSIDrumFabrication}
The drum fabrication process starts with a 2-inch quartz wafer. For each patterning step, we spin-coat three resist layers (MMA(8.5)MAA EL7 at 4000 RPM (rotations per minute) for \SI{60}{\second} with a \SI{90}{\second} bake at \SI{150}{\celsius}, PMMA 950 A3 at 4000 RPM for \SI{60}{\second} with a \SI{90}{\second} bake at \SI{180}{\celsius}, and then Espacer 300Z at 4000 RPM for \SI{60}{\second} with a \SI{60}{\second} bake at \SI{90}{\celsius}) and pattern the design by electron beam lithography. Our development recipe is a \SI{30}{\second} bath in a 1:3 mixture of MIBK:IPA, followed by a brief bath and rinse in pure IPA. The first pattern step consists of markers for alignment, so we evaporate \SI{5}{\nano\meter} Ti and \SI{40}{\nano\meter} Au. Then we perform lift-off using a hot acetone bath ($\simeq 55$~\si{\celsius} for 20 minutes followed by 2 minutes in a sonicator bath).

The bottom layer pattern consists of the microwave cavity, bonding pads, waveguides and the bottom plate of the drum resonator. We evaporate a \SI{40}{\nano\meter} Al layer in an electron beam evaporator, and realise the liftoff as described before. Then, we grow a \SI{80}{\nano\meter} layer of amorphous silicon ($\alpha$-Si) by PECVD, which forms the sacrificial layer between the drum plates. This layer is patterned with a single layer of PMMA 950 A9, and etched by reactive ion etching using a mixture of SF$_6$/O$_2$. 

The top layer consists only of the top drum plate. We evaporate a \SI{120}{\nano\meter} Al layer after patterning. Until now, all steps have been performed on a full 2-inch wafer, but after the final Al evaporation we dice the chips to their final $4 \times 4$~\si{\milli\meter\squared} size. We perform a heat treatment (15 minutes on a hot plate set to between 240 and 280~\si{\celsius}) to redistribute the stress in the top-layer Al and ensure the drums are flat (i.e., not bulging upwards). The final step is the release etch, where we etch the $\alpha$-Si sacrificial layer away with a reactive ion etch mixture of SF$_6$/O$_2$ and get a drum resonator with a suspended top layer.

\subsection*{Measurement setup}\label{SecSIMeasurementsetup}
\begin{figure}
\includegraphics[width = 0.40\textwidth]{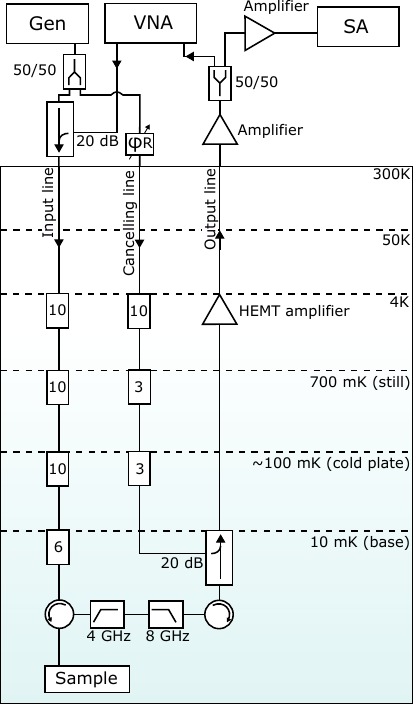}
\caption{\textbf{Setup.} Schematic of the setup used in the experiments. The two drives $a_\mathrm{mw}$ and $a_\mathrm{sb}$ are sourced from a microwave generator (Gen) and a vector network analyzer (VNA) respectively. The output signal is split between the vector network analyzer and a spectrum analyzer (SA). The attenuator values are given in dB.}
\label{FigSetup}
\end{figure}

Our setup consists of a sample mounted on the base plate of a Bluefors dilution refrigerator, as schematically shown in Fig.~\ref{FigSetup}. To drive our sample, we source the drive $a_\mathrm{mw}$ at the cavity frequency from a microwave generator, and the drive at the red sideband, $a_\mathrm{sb}$, from a vector network analyzer. The drive $a_\mathrm{sb}$ is attenuated by a directional coupler that merges it to the main drive $a_\mathrm{mw}$. Both drives make their way through a suitably attenuated input line to the sample at the base plate of the refrigerator. 

The sample is measured in reflection, and the reflected signal is routed through a stack of two circulators and a band-pass filter. A copy of $a_\mathrm{mw}$ split off from the same source is used to interferometrically cancel the drive component reflected from the sample. This is done to avoid saturating the cryogenic amplifier. The cancellation line has a tunable phase delay and attenuation, which we manually set to maximally cancel the output before starting any measurement. On the output line from the sample, after the directional coupler, there is a low-temperature low-noise amplifier, followed by another amplifier at room temperature. Finally, half of the signal is sent back to the network analyzer for measurement, while the other half is routed through a third amplifier to the spectrum analyzer.  

Our measurement protocol starts by defining a set of frequencies around the sideband that we will drive. This gives us control over the direction of the frequency sweep. The cavity drive is turned on, then the sideband drive is initialized to the first frequency. We record the signal on the spectrum analyzer while the sideband frequency is swept. We repeat this sequentially for each of the six readout windows, then reverse the frequency sweep direction, and repeat it again for all six readout windows. This stepped protocol means that the sideband drive is briefly turned off to switch to the next frequency, which can cause the oscillator to decay to the low-amplitude branch. The point on the high-amplitude branch at which this happens is not random, it is reproducible between measurements, as evidenced by the overlap of the curves in Fig.~\ref{FigOptomechanicsCasimir2}\textbf{b}. The stepped drive frequencies do not overlap perfectly with the frequency bins of the spectrum analyzer, and we apply a data filtering scheme detailed in the \ref{SI}.

\subsection*{Drum frequency and effective mass}\label{SecSIDrumSimulation}
\begin{figure}
\includegraphics[width = 0.5\textwidth]{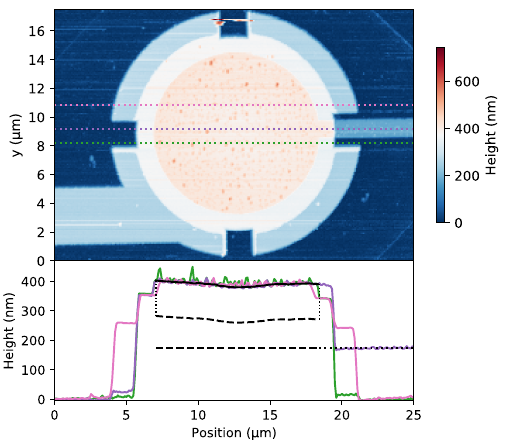}
\caption{\textbf{AFM measurement of the superconducting drum.} The room-temperature gap between the superconducting plates can be estimated from an AFM measurement by comparing the heights along different linecuts (colored dotted lines). The purple line provides a height reference for the bottom plate via the exposed connector on the right side; the green line provides a height reference for the (etched) substrate and the pink line provides a height reference to the thickness of the top layer. We extrapolate the gap (dashed black lines) by subtracting the top layer thickness, \SI{120}{\nano\meter} from the smoothed top linecut. There is a slight  concavity of the suspended part, approximately \SI{15}{\nano\meter} on an average gap of \SI{100}{\nano\meter}.}
\label{FigDrumAFMmeasurement}
\end{figure}

The released drums are measured in an atomic force microscope (AFM) to estimate the gap at room temperature. The device measured in this work is shown in Fig.~\ref{FigDrumAFMmeasurement}. 
By comparing the heights at various points on the geometry of the drum, we can estimate the layer thicknesses and gap size. The connector to the right in Fig.~\ref{FigDrumAFMmeasurement} is deposited in the first evaporation step, so it provides a height reference for the bottom plate of the drum (purple dotted linecut). The top layer shows three distinct heights (along the pink dotted linecut): In the drum center the total thickness is contributed by the thickness of the bottom aluminium plate, the gap and the thickness of the top layer. Closer to the drum edge, the bottom plate stops and only the etched $\alpha$-Si layer and the top layer thicknesses contribute. At the edge of the drum, the top plate contacts the substrate and accounts for the whole thickness. Finally, beyond the drum we measure only the substrate, although the exposed substrate is etched by $\simeq$\SI{120}{\nano\meter} in the drum release step.

\begin{figure}
\includegraphics[width = 0.5\textwidth]{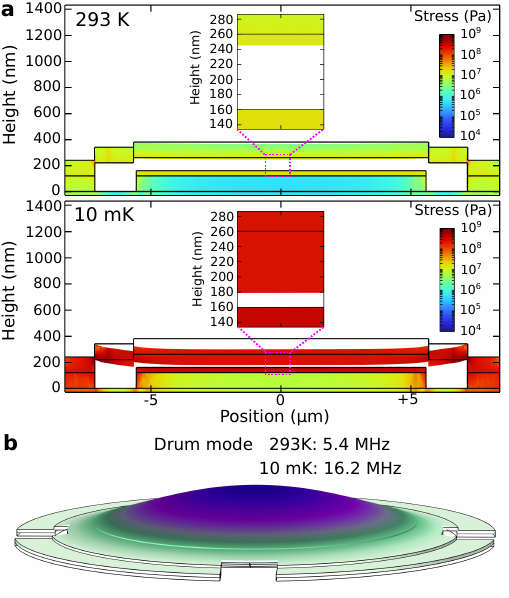}
\caption{\textbf{Simulated drum mechanics. a}: A plane cut through the center of the simulated drum geometry shows a \SI{15}{\nano\meter}  concavity if a \SI{15}{\mega\pascal} tensile pre-stress is included in the aluminium domain. By cooling down to \SI{10}{\milli\kelvin}, the aluminium contracts more than the quartz substrate, which greatly increases the tensile stress in the drum. 
Under this tensile stress, the torque applied on the edges of the drum brings it downwards which reduces the vacuum gap from \SI{100}{\nano\meter} to very low values consistent with the experimentally determined \SI{18}{\nano\meter} according to simulations.
\textbf{b}: Simulated mode shape and frequencies of the drum. The fundamental drum mode greatly increases in frequency due to the added tensile stress. 
}
\label{FigDrumSimulations}
\end{figure}

We smooth the measurement of the top drum layer and extrapolate it using the known layer thickness to calculate the average gap. The extrapolated gap is shown by dashed black lines in Fig.~\ref{FigDrumAFMmeasurement}, and it is approximately \SI{100}{\nano\meter}. "There is a slight concavity in the middle of the drum of \SI{15}{\nano\meter}. This can be explained by a small tensile stress that remains in the aluminium layer due to fabrication. We model the geometry of the drum in COMSOL and show a cut plane through the center of the drum in Fig.~\ref{FigDrumSimulations}\textbf{a}. At room temperature, a \SI{15}{\mega\pascal} tensile stress in the aluminium domain yields a \SI{15}{\nano\meter}  concavity that matches the AFM measurement. 

When the drum is cooled down, the materials thermally contract, but the contraction of the aluminium is much stronger than that of the quartz. To estimate the relative contractions, we use the temperature dependent material parameters included in COMSOL's basic library, for thin-film aluminium and c-axis quartz (corresponding to our z-cut wafers). We extrapolate the dilation coefficient of quartz below \SI{73}{\kelvin} based on literature values~\cite{Barron1982}. 
As a result of the larger relative contraction of the aluminium drum with respect to its substrate, it is subject to an increased tensile stress upon cooling it down to $10\,\rm mK$. The side walls of the drum  unfold and the membrane is brought downwards, which has been previously noted by other groups working with drum resonators~\cite{Andrews2015}. The gap is reduced from \SI{100}{\nano\meter} to \SI{18}{\nano\meter}. 
Note that this estimate of the drum's separation $d = 18$~\si{\nano\meter} (unperturbed by the Casimir force) is found from an entirely independent estimation method from the value of the same parameter $d = 18.00 \pm 0.25$~\si{\nano\meter} found by fitting experimental data. While we consider that this simulation method to estimate $d$ is much less reliable than the fit to experimental data, the fact that these methods independently yield the same value gives confidence in this estimate.

The shape of the fundamental drum mode (Fig.~\ref{FigDrumSimulations}\textbf{b}) is negligibly affected by the thermal contraction, but the frequency of the mode 
is simulated to vary from \SI{5.4}{\mega\hertz} at room temperature to  
\SI{16.2}{\mega\hertz} at \SI{10}{\milli\kelvin}. This frequency is the 'unperturbed' frequency in Table \ref{Tableparameters} that we use as input for the Casimir oscillator simulations.

The mechanical mode shape ${\bf u}({\bf r})$ computed using COMSOL is shown in Fig.~\ref{FigDrumSimulations}\textbf{b}. Here, the displacement detection and actuation profiles coincide and are commonly denoted ${\bf w}({\bf r})$. In this case the effective mass is computed as~\cite{Pinard1999}
\begin{equation}
m_\mathrm{eff} =  m \frac{ \int_V \, || {\bf u}({\bf r})||^2\,  d^3 {\bf r}/V}{  \left( \int_V {\bf w}({\bf r}) \cdot {\bf u}({\bf r})  \, d^3 {\bf r}/V \right)^2},
\end{equation}
where $m$ is the physical mass, $V$ is the volume of the oscillator. The amplitude given to ${\bf u}({\bf r})$ for the computation is irrelevant. In a good approximation, the actuation/detection profile ${\bf w}({\bf r})$
 is assumed to be oriented along $z$-axis, and to be homogeneous across the bottom plate. The physical mass is  $m= 5.4\cdot 10^{-14}$~\si{\kilo\gram} and the effective mass $m_{\rm eff}= 3.96\cdot 10^{-14}$~\si{\kilo\gram}.

\subsection*{Simulations of the dynamics in a Casimir potential}\label{SecSIMatCont}
In Eq.~\eqref{EqCasimirEOM} we have added a strongly nonlinear term to the equation of motion of a harmonic oscillator. It is typical to perform a Taylor expansion of this term for small motions and truncate the resulting series~\cite{Chan2001,Chan2001a}, but that approximation fails to describe the higher-amplitude motion seen in our measurements. Instead, we use a the MATLAB-based continuation library MatCont~\cite{Dhooge2003,Dhooge2008} for our numerical computations, which can handle Eq.~\eqref{EqCasimirEOM} without approximations. This toolbox employs different methods for an extensive study of nonlinear dynamics of a system, such as tracking equilibrium points, detecting bifurcations, and continuing branches of periodic solutions. The continuation method involves two steps: iterative prediction of a point on the solution curve and correction of the predicted point through a Newton-like procedure. This procedure is based on linearizing the function at the current guess and finding where the linear approximation intersects with the x-axis, which becomes the next approximation.

For MatCont, the system of equations must be of the first-order and autonomous (no explicit time dependence). The drive term of Eq.~\eqref{EqCasimirEOM}, $F_0 = F_\mathrm{d} \sin(\omega_\mathrm{d}t)$, is the only term explicitly dependent on time. We can make it autonomous by using the Hopf normal form which allows us to express the dynamics in terms of two real variables in amplitude-phase form. Hopf normal form in Cartesian coordinates reads

\begin{equation}
    \begin{split}
        	& \dot{u} = (\mu - x^2 - y^2)x - \omega_\mathrm{d} y \\
	& \dot{v} = (\mu - x^2 - y^2)y + \omega_\mathrm{d} x 
    \label{eqn:hopf_cart}
    \end{split}
\end{equation}

Assuming $\mu = 1$, we choose $u$ to substitute the drive term.

We non-dimensionalize and scale Eq.~\eqref{EqCasimirEOM}, and split it into two first-order equations. We define,
\begin{equation}
x' = \frac{x}{x_0}, \quad t' = \frac{t}{t_0}, \quad \mathrm{and} \quad y = \dot{x} 
\end{equation}
where $x_0 =  10^{-9}$ m and $t_0 = 2\pi / \omega_\mathrm{r}$.

Substituting all, our system of equations become:
\begin{equation}
\begin{aligned}
\dot{x} &= y, \\
\dot{y} &= -Ay - Bx - C(x+d)^n + Du, \\
\dot{u} &= u(1-u^2-v^2) - \Omega u, \\
\dot{v} &= v(1-u^2-v^2) + \Omega v,
\end{aligned}
\end{equation}
with $A= t_0 \gamma_\mathrm{r}$, $B = t_0^2 \omega_\mathrm{r}^2$, and
\begin{equation}
C = \frac{t_0^2 P_\mathrm{c} \pi r^2 d^n}{m_\mathrm{eff}(x_0)^{n+1}}, \quad \mathrm{and} \quad D = \frac{t_0^2 F_\mathrm{d}}{m_\mathrm{eff} x_0}.
\end{equation}

We apply a periodic forcing to the system by arbitrarily choosing $u = 1$, $v=0$. First, we perform an extended simulation until the system converges to a limit cycle. The last orbit provides the initial conditions for the continuation simulation. By varying certain parameters such as the drive force amplitude $F_d$, damping coefficient $\gamma_r$, separation distance $d$ and the Casimir exponent $n$, we study how the system's response is affected. 

\vspace{0.25cm}
We have become aware of a recent work studying the Casimir force across a superconducting transition~\cite{Xu2025}.

\vspace{0.25cm}
\paragraph*{Data and code availability}\mbox{}\\
All data, simulations, measurement and analysis scripts in this work are available at \href{https://doi.org/10.5281/zenodo.14700381}{https://doi.org/10.5281/zenodo.14700381}.

\vspace{0.25cm}
\paragraph*{Acknowledgments}\mbox{}\\
We would like to acknowledge Gongchang Lin for help with the AFM and Eddy Collin for help with the geometric nonlinearity. We acknowledge the facilities and technical support of Otaniemi research infrastructure for Micro and Nanotechnologies (OtaNano), and the Aalto Scientific Computing team for their support. We also acknowledge the financial support of the Finnish Ministry of Education and Culture through the Quantum Doctoral Education Pilot Program (QDOC VN/3137/2024-OKM-4) and the Research Council of Finland through the Finnish Quantum Flagship project (project number 358877, Aalto University). L.M.d.L. acknowledges funding from the Strategic Research Council at the Academy of Finland (Grant No. 338565).

\vspace{0.25cm}
\paragraph*{Author contributions}\mbox{}\\
M.d.J., E.K., L.B., and L.M.d.L. designed, developed, and performed the experiments. L.B. developed and fabricated the device, L.M.d.L, M.d.J. and E.K. developed the theoretical and numerical analysis. L.M.d.L. conceptualized the project and L.M.d.L. and M.A.S oversaw it. All authors contributed to writing and editing the manuscript

\vspace{0.25cm}
\paragraph*{Competing interest declaration}\mbox{}\\
The authors declare no competing interests.

\clearpage

\setcounter{figure}{0}
\renewcommand{\thefigure}{S\arabic{figure}}
\renewcommand{\theHfigure}{S\arabic{figure}}
\setcounter{equation}{0}
\renewcommand{\theequation}{S\arabic{equation}}
\renewcommand{\theHequation}{S\arabic{equation}}
\section*{Supplementary Information}\customlabel{SI}{Supplementary Information}
This part of the document contains the supplementary information.

\subsection{Casimir and van der Waals regimes}\label{SISecCasimirvdW}

\begin{figure}[h]
\centering
\includegraphics[width = 8cm]{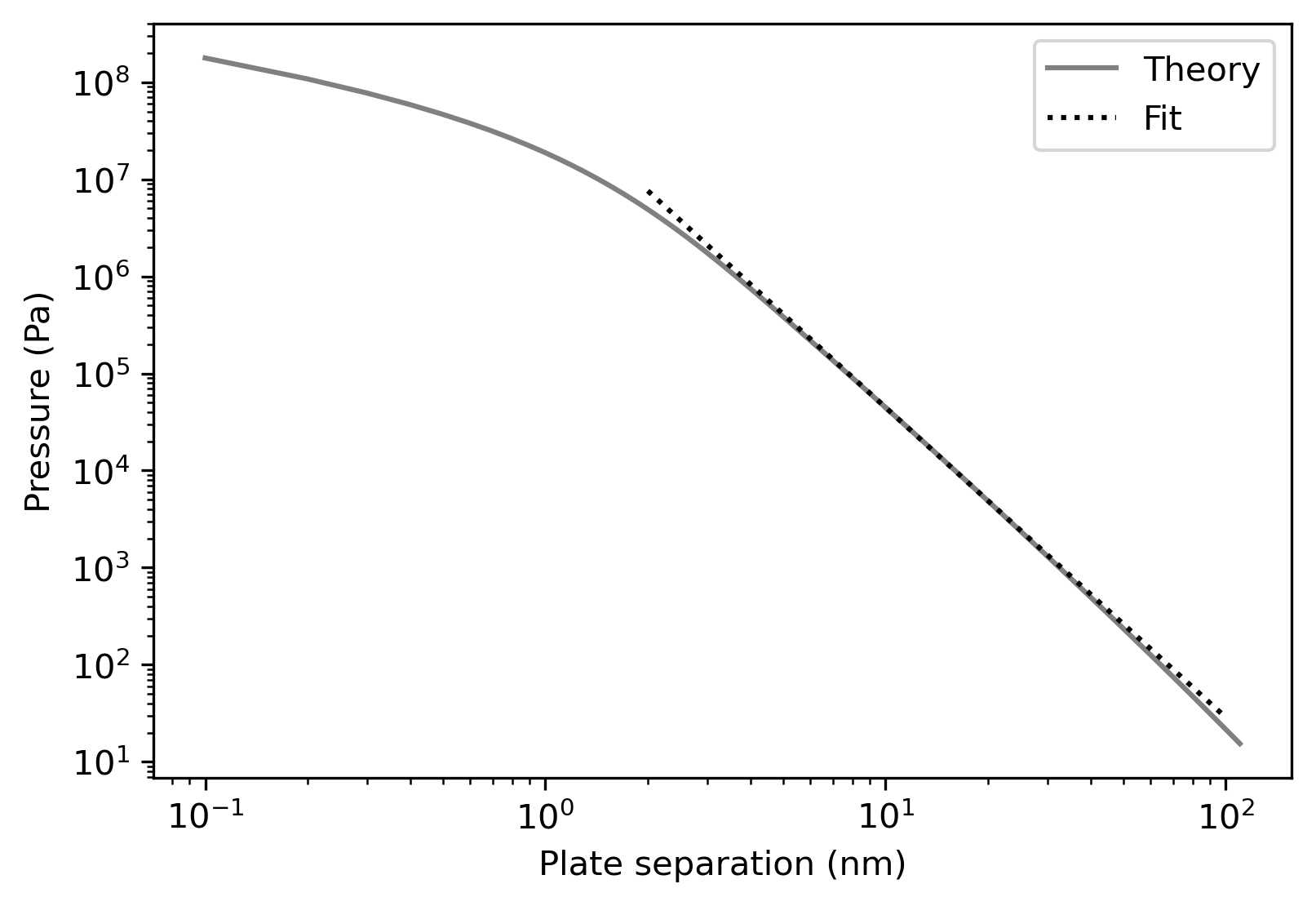}
\caption{Calculation of the pressure across a large range of distances in the Lifshitz formalism. The dotted line shows the Casimir force expression used in the main text, and it is plotted over a large range of separations of the order of magnitude expected in the device $2-100\,\rm nm$.}
\label{FigCasimirvdW}
\end{figure}

Here we justify that the system studied in this work is expected to operate in the force regime commonly referred to as Casimir regime rather than the regime called van der Waals regime. The two regimes are defined for example in Ref.~\cite{Lifshitz1956}, which is a seminal work on the topic of the Casimir force by Lifshitz. In the van der Waals regime, the retardation of the response of one object to charge fluctuations in the other object (or electromagnetic fluctuations at the other object, depending on the point of view adopted to describe the force) is negligible. This regime corresponds to small separations. The Casimir regime corresponds to larger separations where the retardation is important. The Lifshitz model for the estimation of dispersion forces predicts both regimes and allows to define a threshold between the two, as illustrated in Fig.~\ref{FigCasimirvdW}. In our equation of motion for the mechanics, we use the fitted strength and scaling of the Casimir force that is shown by the dotted line in Fig.~\ref{FigCasimirvdW}. The van der Waals regime appears at separations of $1-3\,\rm nm$ nanometers, much smaller than separations considered here.

In Ref.~\cite{Palasantzas2008}, it was shown that the effective threshold of distance between realistic surfaces (e.g., with non-negligible roughness) that separate the van der Waals and the Casimir regimes is of the order of 10\% of the plasma wavelength. For aluminium, this crossover is then expected around \SI{9.5}{\nano\meter}. Although this is closer to the experimental separation estimated from the model of force employed in the paper, it is still inferior, and the van der Waals regime has to be understood as a limit case (for separations much smaller than the threshold). Therefore, in the experimental situation explored in this paper, we refer to the Casimir force.

\subsection{Optomechanical description}\label{SISecOptomechanics}
\subsubsection{Equations of motion}
At small amplitude of displacement around the equilibrium position shifted by the nonlinear potential, the optomechanical equations of motion 
are~\cite{Aspelmeyer2014}:
\begin{equation}
\begin{aligned}
\dot{\tilde{x}}(t) &= \omega_\mathrm{m} {p}, \\
\dot{{p}}(t) &= -\omega_\mathrm{m} \tilde{x} - \gamma_\mathrm{m} {p} - g_0 |a|^2, 
\\
\dot{{a}}(t) &= -(i\Delta + \kappa/2){a} + ig_0 \tilde{x}{a} + \sqrt{\kappa_\mathrm{e}}{s}_\mathrm{in}(t) 
 \\
\label{EQ_SI_EOM_full}
\end{aligned}
\end{equation}
The displacement and momentum are respectively denoted $\tilde{x}$ and ${p}$, 
and ${a}$ denotes the cavity mode field complex amplitude. The mechanical parameters $\omega_\mathrm{m},~\gamma_\mathrm{m}$, the cavity parameters $\kappa,~\Delta$, and the coupling strength $g_0$ are the same as in the main text and their values are listed in Table I. Note that we use the mechanical parameters $\omega_\mathrm{m}$ and $\gamma_\mathrm{m}$ which are the 'dressed' parameters from the Casimir force rather than the 'unperturbed' parameters $\omega_\mathrm{r}$ and $\gamma_\mathrm{r}$. We operate in a frame rotating at the frequency of the microwave drive. 
Our system is strongly driven, so we can safely neglect the noise terms in the 
numerical simulations and use a classical formalism. The input field is a combination of the (strong) drive at the cavity frequency, and a weaker sideband drive. It can be expressed in the rotating frame as
\begin{equation}
{s}_\mathrm{in}(t) = \sqrt{P_\mathrm{mw}/\hbar\omega_\mathrm{c}} + \sqrt{P_\mathrm{sb}/\hbar\omega_\mathrm{c}} e^{-i\omega_\mathrm{m}t}.
\end{equation}
This is enough to solve the system of equations of motion for some initial conditions $\tilde{x}(0)$, ${p}(0)$, and ${a}(0)$. We model the detected signal using  input-output theory 
\begin{equation}
{a}_\mathrm{out} = \sqrt{\kappa_\mathrm{e}} {a} - {s}_\mathrm{in}.
\end{equation}
The spectrum of the output field is given by the Fourier transform $\mathcal{F}$,
\begin{equation}
S_\mathrm{out}(\omega) = \left|\mathcal{F} \left\{ ({a}_\mathrm{out} + {a}_\mathrm{out}^*)/2 \right\}\right|^2.
\label{EqSI_outputfield}
\end{equation}
This spectrum $S_\mathrm{out}(\omega)$ is what we record on the spectrum analyzer.

We can independently calibrate the parameters $\omega_\mathrm{m},~\gamma_\mathrm{m},~\kappa,~\Delta$, and $g_0$ as reported in Sec.~\ref{SecSICharacterization}. We also accurately know the powers of the cavity and sideband drive tones we send in, but between the output of the generators and the cavity are attenuators and cable losses (see \ref{Methods}). To calibrate the amplitudes of our drives at the cavity entrance, we first make a crude estimation based on the relative sideband powers (next section), and then refine it using Eq.~\ref{EqSI_outputfield} (section after that). 

\subsubsection{Relative sideband powers}
\begin{figure*}[t]
\includegraphics[width = \textwidth]{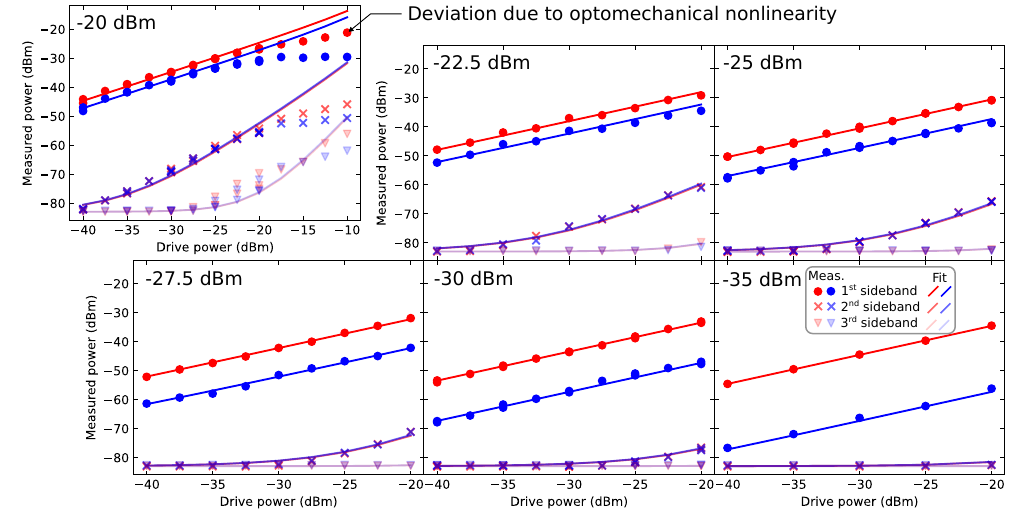}
\caption{\textbf{Relative sideband powers.} Comparison of the maximum powers in all 6 sidebands for different powers of cavity drive $a_\mathrm{mw}$ (panel labels) and sideband drive $a_\mathrm{sb}$ (x-axis). The colors indicate whether the sideband is on the red side of the cavity ($a_{-1,-2,-3}$) or on the blue side ($a_{+1,+2,+3}$). The markers are measurements, solid lines are fits to the model of Eq.~\eqref{EqSidebandamplitudesoutput}. Only at the highest powers does our model deviate from the data, which is due to the optomechanical nonlinearity. The asymmetry of the first-order sidebands provides a power reference between the applied drives. The higher order sidebands are symmetric, so they overlap and appear purple in the plot. This symmetry motivates neglecting the scattering processes of $a_\mathrm{sb}$ which we have done in constructing Eq.~\eqref{EqSidebandmatrix}.}
\label{FigSidebandamplitudes}
\end{figure*}
We have recorded the powers of six sidebands ($\pm \omega_{m}$, $\pm 2 \omega_\mathrm{m}$, and $\pm 3 \omega_\mathrm{m}$) for the measurements reported in the main text. We can compare the relative powers of each of the sidebands, and compare the powers with the sideband drive, to show how our detected signal is proportional to the mechanical displacement. We use the perturbative treatment of the classical coupled mode equations~\cite{Kippenberg2007,Aspelmeyer2014}. Our starting point is a simplified version of Eqs.~(60) and (61) of Ref.~\cite{Aspelmeyer2014}, where the steady state cavity field amplitude $a_0$ is defined by the input power of the microwave drive at the cavity frequency, $a_\mathrm{mw}$, and cavity parameters $\kappa_\mathrm{e}$, $\Delta$ and $\kappa$. We define $x_{\rm amp}$ as a quadrature of the mechanical signal and choose the origin of times such that $\tilde{x}= x_{\rm amp}(t) \sin(\omega_m t)$.
\begin{equation}
\begin{aligned}
a_0 &= a_\mathrm{mw} \frac{\sqrt{\kappa_\mathrm{e}}}{-i\Delta + \kappa/2} \\
a_{+1} &= \frac{g_0 {{x}_{\rm amp}} }{2 x_\mathrm{zpf}} \frac{a_0}{-i(\Delta + \omega_\mathrm{m}) + \kappa/2} \\
a_{-1} &= \frac{g_0 {{x}_{\rm amp}} }{2 x_\mathrm{zpf}} \frac{a_0}{-i(\Delta - \omega_\mathrm{m}) + \kappa/2}.
\label{EqScatterinsideband}
\end{aligned}
\end{equation}
Here, the amplitude of the anti-Stokes- and Stokes-scattered sidebands are denoted with $a_{+1}$ and $a_{-1}$ respectively. These exist in the spectrum at $\pm \omega_\mathrm{m}$ away from the frequency of $a_0$, and they are related to the mechanical amplitude ${x}_{\rm amp}$. This expansion is valid regardless of the spectrum of ${x}_{\rm amp}$, and simply requires that it is peaked close to $0$ frequency and sufficiently narrow so that the sidebands are well-separated.

Our six-sideband treatment is based on repeating the treatment of Eq.~\eqref{EqScatterinsideband}, but considering $a_{\pm 1,2,3}$ as the source term instead of $a_0$. That is, from $a_{+1}$ we recognize two scattering processes, which end up at $a_0$ and $a_{+2}$ respectively. 

\begin{equation}
\begin{aligned}
a_0 &=+ \frac{g_0 {x}_{\rm amp}}{2 x_\mathrm{zpf}} \frac{a_{+1}}{-i\Delta + \kappa/2} \\
a_{+2} &= \frac{g_0 {x}_{\rm amp}}{2 x_\mathrm{zpf}} \frac{a_{+1}}{-i(\Delta + 2 \omega_\mathrm{m}) + \kappa/2}
\end{aligned}
\end{equation}

By repeating this treatment for all orders of sidebands that we measure, we gain a set of linear coupled equations. This resulting system can be written in matrix form as 
\begin{equation}
\begin{bmatrix}
1 & d_{+3} & 0 & 0 & 0 & 0 & 0 \\
d_{+2} & 1 & d_{+2} & 0 & 0 & 0 & 0 \\
0 & d_{+1} & 1 & d_{+1} & 0 & 0 & 0 \\
0 & 0 &d_{0} & 1 & d_{0} & 0 & 0 \\
0 & 0 & 0 & d_{-1} & 1 & d_{-1} & 0 \\
0 & 0 & 0 & 0 & d_{-2} & 1 & d_{-2} \\
0 & 0 & 0 & 0 & 0 & d_{-3} & 1
\end{bmatrix}
\begin{bmatrix}
a_{+3} \\
a_{+2} \\
a_{+1} \\
a_{0} \\
a_{-1} \\
a_{-2} \\
a_{-3}
\end{bmatrix}
= 
\begin{bmatrix}
0 \\
0 \\
0 \\
\frac{a_\mathrm{mw}\sqrt{\kappa_\mathrm{e}}}{-i\Delta + \kappa/2} \\
0 \\
0 \\
0
\end{bmatrix}
\label{EqSISidebandmatrix}
\end{equation}
with
\begin{equation}
d_{\pm n} = \frac{g_0 {x}_{\rm amp}}{2 x_\mathrm{zpf}} \frac{1}{-i(\Delta \pm n\omega_\mathrm{m}) + \kappa/2}.
\end{equation}
We have made the simplifying assumption that repeated interactions ending up at lower-order sideband are negligible in power compared to the (original) lower-order sideband power. This is a valid assumption when $g_0 \ll \kappa/2$, since each individual photon is much more likely to exit the cavity than to scatter from the mechanical resonator. 

It is fairly straightforward to numerically solve Eq.~\eqref{EqSISidebandmatrix} using e.g., Python. This model is valid when the optomechanical interaction is dominated by a single drive tone, which in our case is $a_\mathrm{mw}$. However, we have a weaker second drive, $a_\mathrm{sb}$, at $-\omega_\mathrm{m}$. Since it is much weaker than $a_\mathrm{mw}$, we neglect the effect of any scattering interactions from $a_\mathrm{sb}$ on the sideband amplitudes. The result of this is that there is an asymmetry only between the first-order red ($a_{-1}$) and blue ($a_{+1}$) sidebands. If we had included the scattering interactions of the weaker drive, the red and blue sidebands of all orders would have the same asymmetry. However, from the measurements we will see that this is not the case (only the first order sidebands are asymmetric), and therefore we neglect the scattering interactions of $a_\mathrm{sb}$ in our sideband amplitude calculation. Thus the resulting amplitude(s) due to this drive are
\begin{equation}
\vec{a}_\mathrm{sb} = 
\begin{bmatrix}
0 \\ 0 \\ 0 \\ 0 \\ \frac{a_\mathrm{sb} \sqrt{\kappa_\mathrm{e}}}{(-i(\Delta - \omega_\mathrm{m}) + \kappa/2)} \\ 0 \\ 0
\end{bmatrix}.
\end{equation}

To fit this model to our data, we solve Eq.~\eqref{EqSISidebandmatrix} for $\vec{a}$ and compute the output field using the input-output formalism. Here we also add the solution for the weaker drive field $a_\mathrm{sb}$. The output field is described by
\begin{equation}
a_\mathrm{out} = \sqrt{\kappa_\mathrm{e}}(\vec{a} + \vec{a}_\mathrm{sb}) - 
\begin{bmatrix}
0 \\ 0 \\ 0 \\ \frac{a_\mathrm{mw}}{-i\Delta + \kappa/2} \\ \frac{a_\mathrm{sb}}{-i(\Delta - \omega_\mathrm{m}) + \kappa/2} \\ 0 \\ 0
\end{bmatrix}
\label{EqSidebandamplitudesoutput}
\end{equation}

In Fig.~\ref{FigSidebandamplitudes} we have plotted the maximum observed powers of each of our measured sidebands (markers) for different drive powers, and compared those to our fitted model (solid lines). There is an excellent agreement at all but the highest powers. For the highest powers, the optomechanical nonlinearity reduces the drive efficiency, which reduces the maximum powers seen in the sidebands. 

The measurements record the sideband powers, and from our model we extract $|a_\mathrm{out}|^2$, but we need two conversion factors to link the theory to our experiments (one for each axis of Fig.~\ref{FigSidebandamplitudes}). There is also attenuation from the room-temperature source to the cavity at \SI{10}{\milli\kelvin} (see \ref{Methods}), which gives us an additional conversion factor. The simplifications of the model of Eq.~\eqref{EqSISidebandmatrix} are not ideal to use it as an accurate calibration tool between mechanical amplitude and measured power, but it serves as a quick estimation of these conversion factors. We use these as an initial guess for the simulation-based calibration method using the optomechanical equations of motion described in the next section.

\subsubsection{Absolute amplitude calibration}

Here we describe the calibration of measured signals into displacement. After the calibration of optomechanical parameters described in Sec.~\ref{SecSICharacterization}, a remaining unknown is the total attenuation between microwave sources and the cavity entrance. Indeed, input lines include attenuators with known attenuation (see Fig.~\ref{FigSetup}) but also losses incurred along the cables whose exact value is unknown. To realize this calibration, we compare the measured response of the system at small displacements with numerical predictions.

\begin{figure}
\includegraphics[width = 0.48\textwidth]{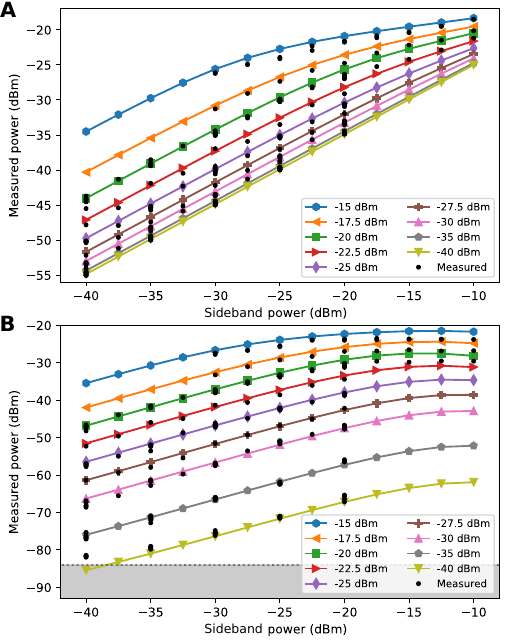}
\caption{\textbf{Sideband calibration. A} Simulated powers (colored lines with various markers) and observed peak powers (black dots) for the red sideband. The legend indicates $P_\mathrm{mw}$, the x-axis is $P_\mathrm{sb}$. The agreement between the simulations and the measurements at low power serves as a calibration for the numerical simulations. \textbf{B} Idem, but for the blue sideband. The grey shaded area shows the noise floor for the measurements.}
\label{FigSI_absoluteamplitudecalibration}
\end{figure}

\begin{figure}
\includegraphics[width = 0.48\textwidth]{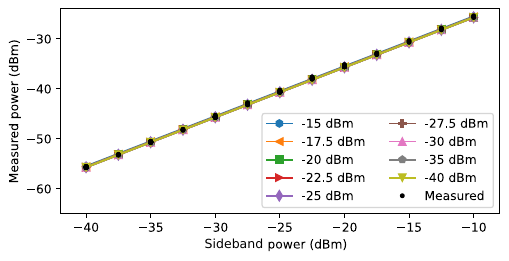}
\caption{\textbf{Input power calibration.} The \SI{3}{\kilo\hertz} off-resonant drive simulations show a perfect overlap between all simulated (colored markers) and measured powers (black dots).}
\label{FigSI_absoluteamplitudecalibration2}
\end{figure}

To model the response, we let Eq.~\eqref{EQ_SI_EOM_full} evolve numerically until it has settled into a steady state, and simulate the steady state for an additional \SI{10}{\milli\second}. We compute the Fourier transform of the output field, as in Eq.~\eqref{EqSI_outputfield}, which gives us the observed power, while we obtain the mechanical oscillation amplitude from the size of the orbit of $\tilde{x}$ 
of Eq.~\eqref{EQ_SI_EOM_full}. 
To scale the simulated signal in the same way as the measured signal, we multiply the simulation result by the detection scale factor (W/photon) $S_{\rm c}$ found in Sec.~\ref{SecSICharacterization} and compensate for the finite simulation time.
We then compare the maximum simulated power around the first red and the first blue sideband ($\omega_c\pm \omega_m$) with the maximum measured amplitude. We calibrate the cables attenuation by matching the computed and measured signals simultaneously for these two sidebands' maxima (see \ref{FigSI_absoluteamplitudecalibration}) as well as for a point on the side of the red sideband detuned by 3 kHz (see Fig.~\ref{FigSI_absoluteamplitudecalibration2}).
Fig.~\ref{FigSI_absoluteamplitudecalibration} shows that the calibrated values allow to match data and simulations for all combinations where both powers are in the range $[{-40\,\rm dB}, -30\,\rm dB]$: these powers correspond to those for which the mechanical responses were measured to be approximately Lorentzian and therefore reasonably well described by  Eq.~\eqref{EQ_SI_EOM_full}. Additionally, \ref{FigSI_absoluteamplitudecalibration2} shows that the same calibrated values allow to match data and simulations at all powers for the off-resonant point where the displacement is also very small.

From this procedure, we find that the power of the cavity drive $P_\mathrm{mw}$ at the input of the device is attenuated by $47.4$~dB from the power issued by the generator. Out of this attenuation, $36$~dB comes from the installed attenuators shown in Fig.~\ref{FigSetup}, $3\,\rm dB$ comes from a splitter at room temperature, and the rest (about $8\,\rm dB$) is of the order of magnitude of the total attenuation expected from the cryogenic transmission line itself. Similarly, we find that the sideband tone power $P_\mathrm{sb}$ at the input of the cavity is attenuated by $69.6$~dB from the power issued by the generator. The difference of $22.2\,\rm dB$ between the attenuations of the two tones is mainly due to the $20\,\rm dB$ attenuation of a directional coupler in front of the vector network analyzer generating the sideband drive (see Fig.~\ref{FigSetup}).

Fig.~\ref{FigSI_absoluteamplitudecalibration} also shows the maximum amplitude of sidebands computed at larger drives. As shown in the figure, while Eq.~\eqref{EqOptomechanicalEOM} cannot reproduce the asymmetric mechanical lineshapes measured at larger drives, its solution still predicts the \textit{maximum amplitude} of measured sidebands rather accurately up to very large driving forces.

\begin{figure}
\includegraphics[width = 0.48\textwidth]{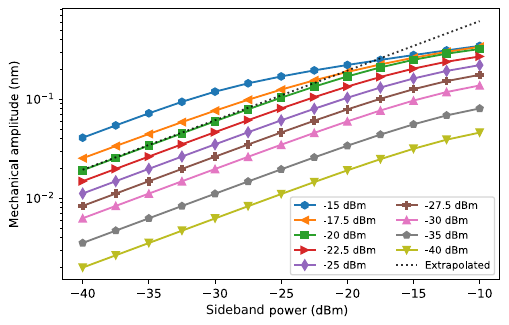}
\caption{\textbf{Mechanical amplitudes simulated with \ref{EqOptomechanicalEOM}} for different combinations of cavity and sideband drive powers.}
\label{FigSI_absoluteamplitudecalibration3}
\end{figure}

The displacement amplitude simulated for all combinations of cavity and sideband drive powers are shown in Fig.~\ref{FigSI_absoluteamplitudecalibration3}. At small amplitudes, the maximum amplitude scales linearly with both cavity and sideband drive powers, but at some point this relation breaks down. This is not due to the nonlinearity of the oscillator since Eq.~(\ref{EqOptomechanicalEOM}) does not simulate it. Instead, this is due to the loss of drive efficiency due to optomechanical effects.

\subsubsection{Calculation of the drive efficiency}
The tenet of (dispersive) cavity optomechanics is that the cavity frequency $\omega_\mathrm{c}$ shifts as a result of the displacement $\tilde{x}$~\cite{Aspelmeyer2014}. In this work, the amplitude of $\tilde{x}$ is significant, and the cavity frequency shift $\frac{\partial \omega_\mathrm{c}}{\partial \tilde{x}}$ causes a mismatch between the cavity frequency and the drive frequencies, $\omega_\mathrm{mw} \simeq \omega_\mathrm{c}$ and $\omega_\mathrm{sb} \simeq \omega_\mathrm{c} - \omega_\mathrm{m}$. The cavity frequency $\omega_\mathrm{c}(t)$ is thus a function of time, it oscillates at $\omega_\mathrm{m}$ depending on the exact trajectory of the drum.

To put it simply, for larger oscillations of $\tilde{x}$, the frequency mismatch between cavity and drive is greater and the power in the cavity decreases. The oscillation is fast with respect to the cavity linewidth, as we are in the resolved sideband limit $\kappa \ll \omega_\mathrm{m}$, which means that the cavity amplitude changes slowly with respect to the oscillation of $\omega_\mathrm{m}$. Furthermore, we also have two drives separated in frequency. So while the power resulting from the drive at $\omega_\mathrm{c}$ is maximal for small amplitudes of $\tilde{x}$, the power resulting from the drive at $\omega_\mathrm{c} - \omega_\mathrm{m}$ peaks for some specific oscillation amplitude of $\tilde{x}$.

We calculate the drive efficiency by solving Eq.~\eqref{EQ_SI_EOM_full} 
with a fixed mechanical amplitude $x$. We let the system simulate only for a brief time, such that the cavity amplitude $a$ has had time to stabilize.
From the last $20$ mechanical periods, we extract the power in the cavity. This brief simulation is repeated for $101$ values of $x$ logarithmically spaced between \SI{1}{\pico\meter} and \SI{1}{\nano\meter} for all different cavity drive powers used in the experiments. 

In the regime that is relevant for our experiment, the drive efficiency is $1$ for small mechanical amplitudes ($<100$~\si{\pico\meter}), and decreases quickly towards zero beyond that. However, at even larger mechanical amplitudes, beyond what we achieve in this work, the drive efficiency recovers to a non-negligible number in a series of bands that represent stable orbits as described in Ref.~\cite{He2020}. 

To fit a measured response driven by a given combination of powers, we first  estimate the driving force $F_0$ (unaffected by the loss of drive efficiency) corresponding to that combination using the calibration detailed in the previous section. We simulate the corresponding response using MatCont as described in \ref{Methods}, then correct it point-by-point to account for the loss of drive efficiency. This procedure is repeated for different values of parameter $d$ appearing in the Casimir force model in order to realize the fit using a standard least-square routine.

\subsection{Simulating the Casimir nonlinearity}\label{SecSICasimirSims}
\begin{figure}
\includegraphics[width = 0.5\textwidth]{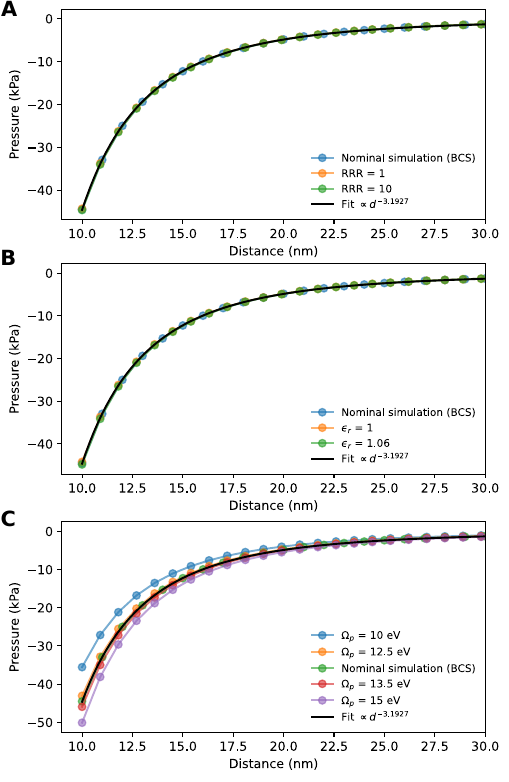}
\caption{\textbf{Effect of material parameters. A:} The residual resistance ratio ($\mathrm{RRR}$) rescales the dissipation rate $\gamma_\mathrm{p} = \gamma_0/\mathrm{RRR}$, but varying it from the nominal value $\mathrm{RRR}=2$ does not significantly affect the Casimir pressure. \textbf{B:} Varying the relative permittivity of aluminium from the the nominal value $\epsilon_\mathrm{r} = 1.03$ does not significantly affect the Casimir pressure. \textbf{C:} Varying the plasma frequency from the nominal value $\Omega_\mathrm{p} = 13$~\si{\eV} affect the Casimir pressure amplitude $P$ more strongly than the Casimir scaling $n$.}
\label{FigCasimirnscaling}
\end{figure}
\subsubsection{Role of Casimir parameters}
We perform the fit of distance $d$ without considering variations in the calculated Casimir pressure, i.e., we consider the Casimir pressure amplitude $P$ and scaling $n$ to be fixed. To motivate this choice, we show that the Casimir pressure does not vary much with the material parameters $\gamma_0$ (or $\mathrm{RRR}$), $\epsilon_0$, and $\Omega_\mathrm{p}$. We repeat the Casimir pressure calculations for different material parameters, as shown in Fig.~\ref{FigCasimirnscaling}, and we fit curves of the form $P_\mathrm{c} = P/d^n$ to distinguish changes in the amplitude of the Casimir pressure ($P$) from changes in the Casimir scaling ($n$).

The residual resistivity ratio $\mathrm{RRR}$ can vary depending on the material source and deposition method. We have previously measured similar films to have $\mathrm{RRR} = 2$ using a four-point probe method. Variations in the $\mathrm{RRR}$ affect the phenomenological dissipation rate, $\gamma_\mathrm{p} = \gamma_0/\mathrm{RRR}$, so Fig.~\ref{FigCasimirnscaling}\textbf{A} shows the effect of varying either $\mathrm{RRR}$ or $\gamma_0$. From the overlap between the curves, we conclude that $\mathrm{RRR}$ and $\gamma_0$ affect neither the Casimir pressure amplitude $P$ nor exponent $n$.

The relative permittivity of aluminium, $\epsilon_\mathrm{r} = 1.03$ was experimentally determined and follows from the core-interband transitions of the aluminium atoms~\cite{Palik1997}. We show in Fig.~\ref{FigCasimirnscaling} that variations in this parameter do not lead to significant changes in the Casimir pressure.

We take the plasma frequency of aluminium $\Omega_\mathrm{p} = 13$~\si{\eV} from Ref.~\cite{Palik1997}. In Fig.~\ref{FigCasimirnscaling} we show that varying $\Omega_\mathrm{p}$ changes the Casimir pressure, but its effect on the Casimir scaling is marginal. A $25\%$ variation in $\Omega_\mathrm{p}$ corresponds to a $25\%$ variation in $P$, but less than $1\%$ variation in $n$.

\begin{figure}
\includegraphics[width = 0.5\textwidth]{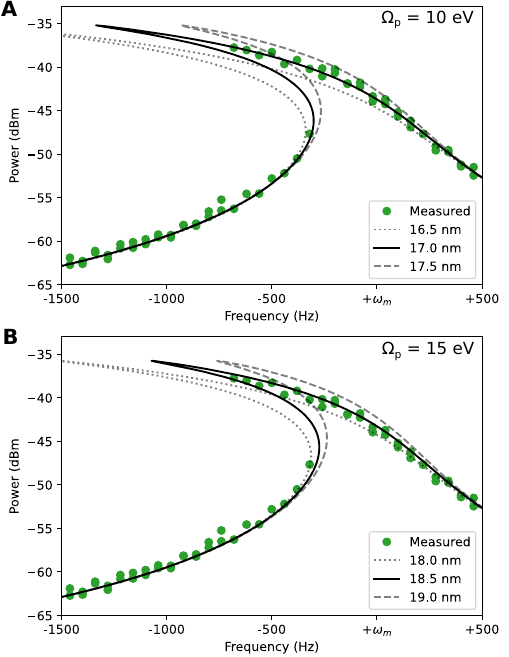}
\caption{\textbf{Fits for Casimir force variations. A:} The Casimir force for a material with plasma frequency $\omega_\mathrm{p} = 10$~\si{\eV} is weaker than the nominal Casimir force used in this work. The Casimir nonlinearity fits to a plate separation of $d = 17.00 \pm 0.25$~\si{\nano\meter}. \textbf{B:} The Casimir force for a material with plasma frequency $\omega_\mathrm{p} = 15$~\si{\eV} fits to a plate separation of $d = 18.50 \pm 0.25$~\si{\nano\meter}.}
\label{FigAlternatefits}
\end{figure}

To further motivate our choice to consider the calculated Casimir parameters as fixed values, we fit our experimental data to the Casimir force calculated with $\Omega_\mathrm{p} = 10$ and $15$~\si{\eV}. The results are shown in Fig.~\ref{FigAlternatefits}, where we have taken the optomechanical calibration of mechanical amplitude to be fixed. Taking $\Omega_\mathrm{p} = 10$~\si{\eV} (Fig.~\ref{FigAlternatefits}\textbf{A}), the Casimir force is weaker, which means that to experience the same nonlinearity at fixed mechanical amplitude, the plate separation $d$ must be reduced. Vice versa, taking $\Omega_\mathrm{p} = 15$~\si{\eV} (Fig.~\ref{FigAlternatefits}\textbf{B}) leads to a better fit at larger $d$. The variations in $\Omega_\mathrm{p}$ are significant ($25\%$ of the total value), and well beyond the uncertainty of which the plasma frequency is established in aluminium, yet the effect on the fitted parameter $d$ is limited ($\sim$\SI{1}{\nano\meter}). Thus, we conclude that our choice to take the Casimir pressure amplitude and scaling to be fixed values leads to a smaller error than the optomechanical amplitude calibration does.

\subsubsection{Static shift of frequency and position}
\begin{figure}
\includegraphics[width = 0.48\textwidth]{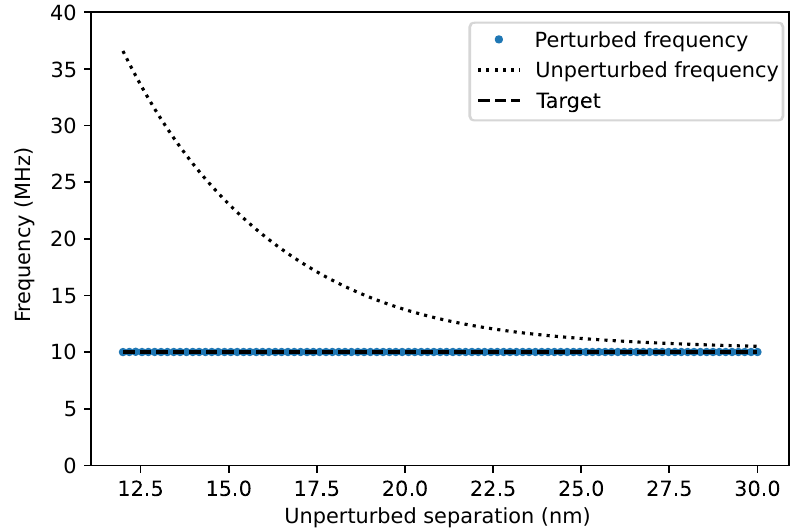}
\caption{\textbf{Casimir frequency shift.} Plot of the unperturbed frequencies $\omega_\mathrm{r}$ for various values of spacing $d$ that result in the correct value for $\omega_\mathrm{m}$ when the Casimir force is included.}
\label{FigSIFrequencyshift}
\end{figure}

As we illustrated in Fig.~\ref{FigSchematicCasimir}\textbf{A}, we start with a harmonic oscillator centered at distance $d$ from the other plate with unperturbed resonance frequency $\omega_\mathrm{r}$. The Casimir force pulls the top plate closer to the bottom plate, such that it oscillates around some value $d' < d$, and softens the spring so the mechanical frequency is decreased, $\omega_\mathrm{m} < \omega_\mathrm{r}$. We do not know $d$ or $\omega_\mathrm{r}$ a priori, but we can measure $\omega_\mathrm{m}$ with great accuracy, leaving us to consider $d$ (or somewhat equivalently the Casimir pressure $P_\mathrm{c}$) as a parameter to be fitted. That means that for every value $d$, we have to find the value of the unperturbed frequency $\omega_\mathrm{r}$ such that the perturbed frequency $\omega_\mathrm{m} = 2\pi \times 10.001$~\si{\mega\hertz}. Rather than finding a clever algorithm to compute the correct values of $\omega_\mathrm{r}$, we  numerically evaluate the problem for a selection of $d$-values and interpolate. In Fig.~\ref{FigSIFrequencyshift} we show the frequencies $\omega_\mathrm{r}$ that result in the correct value of $\omega_\mathrm{m}$ for different values of $d$.

\begin{figure}
\includegraphics[width = 0.48\textwidth]{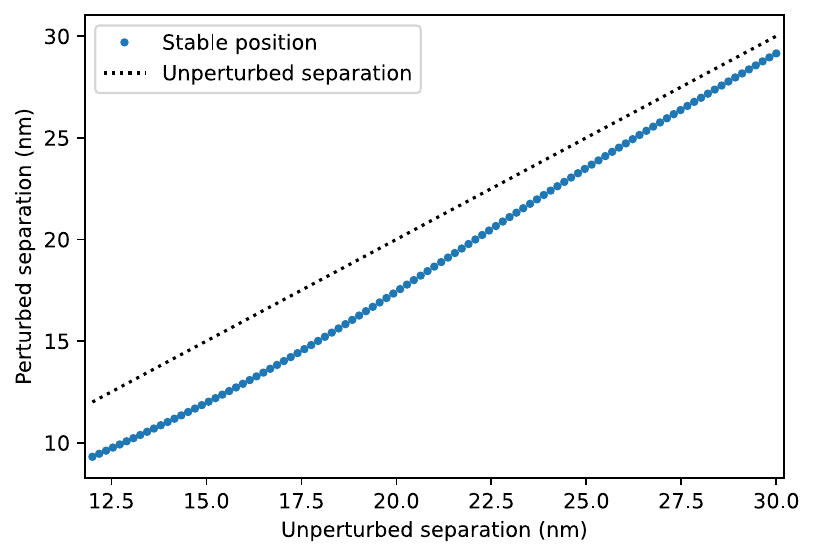}
\caption{\textbf{Shift of the local potential minimum due to the Casimir force.} Distance between the local minimum (stable position) and the bottom plate for various values of the unperturbed separation $d$. The dotted line indicates where the initial separation would be the stable position. For $d = 18.0$~\si{\nano\meter}, the difference between the unperturbed separation and the local minimum is \SI{2.9}{\nano\meter}.}
\label{FigSIPositionshift}
\end{figure}

The numerical simulations of the Casimir oscillator in MatCont start with the drive frequency far detuned. To start the continuation, we need this step to converge to a solution. However, the total potential sketched in Fig.~\ref{FigSchematicCasimir}\textbf{A} has a local minimum (where our resonator is) and a global minimum beyond the pull-in point. We need to feed MatCont the correct initial values such that it converges to the local minimum instead of showing us the pull-in collapse. To find these, we compute the position of the local minimum of the total potential for various values of $d$, while using the values of $\omega_\mathrm{r}$ from Fig.~\ref{FigSIFrequencyshift}. The unperturbed frequency $\omega_\mathrm{r}$ is related to the mechanical stiffness needed to compute the potential, not $\omega_\mathrm{m}$. We plot the stable position of the local minimum for different values of $d$ in Fig.~\ref{FigSIPositionshift}. As with the unperturbed frequencies, we interpolate linearly between the calculated values for the numerical simulations and fits of the main text.

\subsubsection{Role of mechanical parameters}\label{sec:casparameters}
\begin{figure}
\includegraphics[width = 0.5\textwidth]{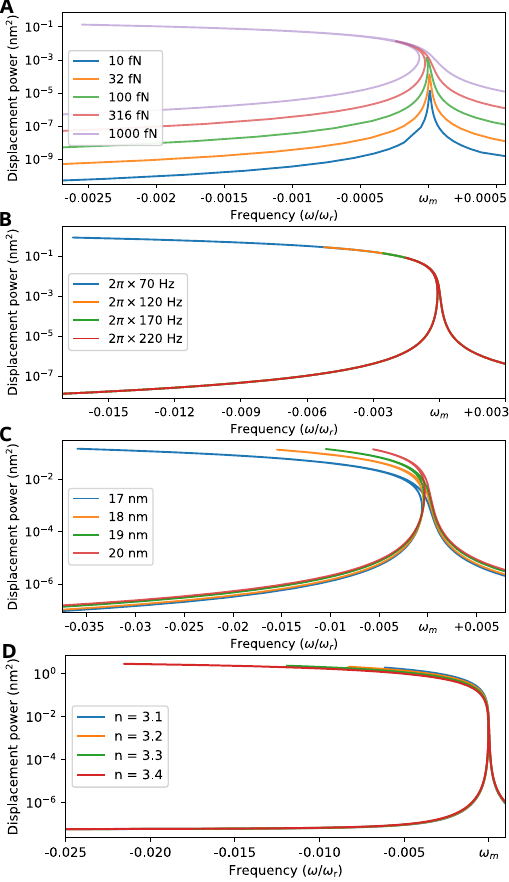}
\caption{\textbf{Effect of the Casimir force parameters.} \textbf{A}: Effect of different drive force amplitudes $F_\mathrm{d}$. \textbf{B}: Effect of varying the damping coefficient $\gamma_\mathrm{r}$. \textbf{C}: Effect of varying the distance between the plates at mechanical zero. $d$. The frequencies of the curves have been shifted to align on $\omega_\mathrm{m}$. \textbf{D}: Effect of varying the Casimir scaling $n$, where the amplitude of the Casimir pressure has been fixed at the equilibrium position $d'$.}
\label{FigCasParameters}
\end{figure}

We study the effect of variations of the parameters of the Casimir effect through numerical simulations. We vary the drive force amplitude $F_\mathrm{d}$, damping coefficient $\gamma_\mathrm{r}$, separation distance $d$, and the Casimir exponent $n$. The results are shown in Figure \ref{FigCasParameters}.

We show how the system response transforms from linear behavior at small amplitude to nonlinear behavior at large amplitude in Fig.~\ref{FigCasParameters}\textbf{A}. When the drive force amplitude is small, $F_\mathrm{d} = 10$~\si{\femto\newton}, the frequency response of the system is Lorentzian, as expected. For larger amplitudes, the resonance peak broadens and becomes asymmetrical, indicative of nonlinear resonance where the response of the system is no longer linearly proportional to the drive force. The resonance frequency decreases with increasing amplitude, which indicates that the Casimir force is a strong softening nonlinearity.

In Fig.~\ref{FigCasParameters}\textbf{B}, we study the effect of the mechanical damping rate $\gamma_\mathrm{r}$. Lower damping allows for greater energy accumulation within the system at equal drive amplitudes, leading to larger oscillations and a more pronounced nonlinearity. However, all the backbones of the curves for different damping rates align, meaning the damping coefficient does not have a significant effect on the nonlinearity we observe.

The most important parameter for Casimir experiments is the separation distance $d$, which is the distance between the plates in the absence of the Casimir force. Since the Casimir effect strongly scales with the distance $d$, the Casimir force between drums spaced $d = 17$~\si{\nano\meter} apart (blue curve in Fig.~\ref{FigCasParameters}\textbf{C}) causes a much stronger nonlinear behavior than for drums spaced $d = 18$~\si{\nano\meter} apart. The region where there are multiple solutions is much more extended, even if the amplitude of the mechanical oscillations remains the same. All curves were generated using the same drive force, which indicates that the separation distance $d$ does not affect the mechanical amplitude significantly. 

Finally, we show the effect of the Casimir force exponent $n$ in Fig.~\ref{FigCasParameters}\textbf{D}. The The Casimir pressure was adjusted for each $n$ such that the value at the equilibrium position ($d'$) was kept constant. While the trend from the change in $n$ somewhat resembles the trend from the change in $d$, it is noticeable that the peaks curve more sharply with increasing $n$, while they start curving at lower amplitude for lower $d$.

\subsection{Characterization of optomechanical system}\label{SecSICharacterization}
\subsubsection{Calibration of microwave cavity parameters}
\begin{figure}
\includegraphics[width = 0.48\textwidth]{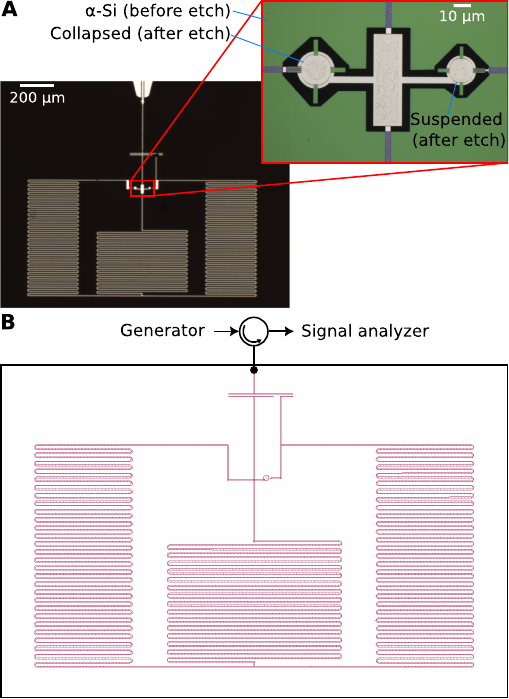}
\caption{\textbf{Microwave cavity. A:} Microscope image of our microwave cavity (after the experiment), with the inset showing the drums (during fabrication). The larger (left) drum collapsed before the experiment, while the smaller (right) drum survived. \textbf{B:} Sonnet simulation of our cavity, with the collapsed drum replaced by a short.}
\label{FigSonnetSimulation}
\end{figure}
Our microwave cavity design was optimized to allow coupling to two separate drum resonators with two cavity modes, as shown in Fig.~\ref{FigSonnetSimulation}\textbf{A}, and described in more detail in the Supplementary of Ref.~\cite{Mercierdelepinay2021}. In our device, one of the drum resonators inadvertently collapsed and shorted one half of the cavity, though this cannot be seen in the microscope image. Instead, we deduce this by the number of microwave modes we see (one), its frequency ($\omega_\mathrm{c} = 2\pi \times 5.4618$0~\si{\giga\hertz}), and the number of mechanical signals we can observe in a wide-band frequency sweep at high red-sideband drive power (one).

We simulate our cavity using Sonnet, where we have replaced the collapsed drum by a short (Fig.~\ref{FigSonnetSimulation}\textbf{B}. At first, we replace the suspended drum by a numerical capacitance. An initial guess for the capacitance of this drum based on the diameter $2r = 11.3$~\si{\micro\meter} and a gap of \SI{15.1}{\nano\meter} is \SI{0.059}{\pico\farad}. To provide an initial estimate of the optomechanical coupling, we vary the spacing of the vacuum layer in our Sonnet model, and re-compute the cavity resonance frequency. We find $\frac{\Delta\omega_\mathrm{c}}{\Delta x} \simeq 2\pi\times 104$~\si{\mega\hertz\per\nano\meter}. We use the zero point motion $x_\mathrm{zpf} = \sqrt{\frac{\hbar}{2m_\mathrm{eff}\omega_\mathrm{m}}} = 4.6$~\si{\femto\meter} and estimate the coupling rate $g_0 \simeq x_\mathrm{xpf} \frac{\Delta \omega_\mathrm{c}}{\Delta x} \simeq 2\pi\times 480$~\si{\hertz}. This is larger than the coupling rate that we find when we calibrate our system by about a factor $3$, which we attribute to details not captured by our simulation: microscopic defects or fabrication imperfections, wire bonds, and box resonances. The simulated microwave cavity mode is not qualitatively affected by these limitations, but the we should treat the optomechanical coupling it yields as an optimistic estimate.

\begin{figure}
\includegraphics[width = 0.48\textwidth]{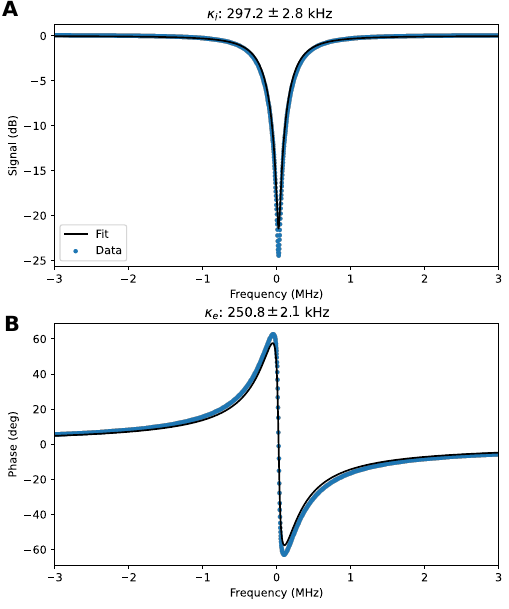}
\caption{\textbf{Microwave cavity calibration. A}: Amplitude response of the microwave cavity (blue) and the fitted linewidth (black). \textbf{B}: Phase response of the microwave cavity (blue) and the fitted response (black). The internal and external linewidths, $\kappa_\mathrm{i} = 297.2 \pm 2.8$~\si{\kilo\hertz} and $\kappa_\mathrm{e} = 250.8 \pm 2.1$~\si{\kilo\hertz} respectively, are fitted as described in the text.}
\label{FigMicrowavecalibration}
\end{figure}
We characterize our microwave resonator by recording its reflected response (both amplitude and phase) and fitting this with the equation~\cite{Aspelmeyer2014}
\begin{equation}
R = \frac{(\kappa_\mathrm{i} - \kappa_\mathrm{e})/2 - i(\Delta - \omega_\mathrm{c})}{(\kappa_\mathrm{i} + \kappa_\mathrm{e})/2 - i(\Delta - \omega_\mathrm{c})}.
\label{Eqcavityresponse}
\end{equation}
Here, the square of the reflection coefficient, $|R|^2$, describes the probability that a photon reflects off our cavity. Furthermore, the internal and external linewidths, $\kappa_\mathrm{i}$ and $\kappa_\mathrm{e}$, sum up to the total linewidth $\kappa = \kappa_\mathrm{i} + \kappa_\mathrm{e}$, and the detuning $\Delta = \omega_\mathrm{d} - \omega_\mathrm{c}$ describes the difference between the drive frequency $\omega_\mathrm{d}$ and the cavity center frequency $\omega_\mathrm{c}$.

We use a three-step fit process. First, we de-trend the cavity signal by fitting a polynomial of order 2 to the first and last 10\% of the amplitude signal, and a polynomial of order 1 to the same parts of the phase signal. This assumes the cavity resonance is approximately at the center of our measurement span, and that the background is reasonably flat within this span. Then we fit an ellipse to the response $R$ on a Smith chart, from which we compute the cavity parameters and use those as initial guesses for the third and final step. By using scipy's \verb|curve_fit| function~\cite{Virtanen2020}, we fit our data to Eq.~\eqref{Eqcavityresponse}. The response, fit and extracted parameters are shown in Fig.~\ref{FigMicrowavecalibration}. We fit the amplitude in logarithmic scale to enhance the accuracy around the cavity center, while the phase is fitted in linear scale. 

Our cavity resonance frequency $\omega_\mathrm{c} = 2\pi \times 5.4618$0~\si{\giga\hertz}, and the internal and external linewidths are $\kappa_\mathrm{i} = 2\pi \times 297.2 \pm 2.8$~\si{\kilo\hertz} and $\kappa_\mathrm{e} = 2\pi \times 250.8 \pm 2.1$~\si{\kilo\hertz} respectively.

\subsubsection{Thermalization of drum motion}
We calibrate the temperature to which the mode of our mechanical resonator thermalizes by sweeping the temperature of the dilution refrigerator. At each temperature, we measure the mechanical spectrum and fit a Lorentzian to the data. This way, we can extract the frequency $\omega_\mathrm{m}$, linewidth $\gamma_\mathrm{m}$, and the area of the mechanical peak.

\begin{figure}
\includegraphics[width = 0.48\textwidth]{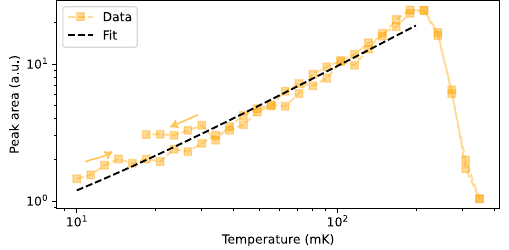}
\caption{\textbf{Thermalization of the drum mode.} Area of the fitted mechanical peaks, indicating the thermal energy of the oscillator mode. The dashed curve is a linear fit offset by only about $3\,\rm mK$, well below the base temperature of $10\,\rm mK$, showing the good thermalization of the mode with the cryostat at the base temperature. At higher temperatures, the microwave cavity response disappears, likely due to quasiparticles in the superconductor (Aluminium), which shifts the microwave cavity frequency.}
\label{FigThermalcalibration}
\end{figure}

The amplitude of the thermal motion peak is shown in Fig.~\ref{FigThermalcalibration}. It increases linearly with temperature  below ~\SI{200}{\milli\kelvin}.  At higher temperatures the quasiparticles in the Al superconducting cavity shift and damp the cavity mode which reduce the readout efficiency. 
The peak amplitude stays proportional to the temperature all the way to the base temperature of \SI{10}{\milli\kelvin}: we conclude that the drum thermalizes down to \SI{10}{\milli\kelvin} and we use this assumption to extract $g_0$ in the next section.

\subsubsection{Calibration of single-photon coupling}
\begin{figure}
\includegraphics[width = 0.48\textwidth]{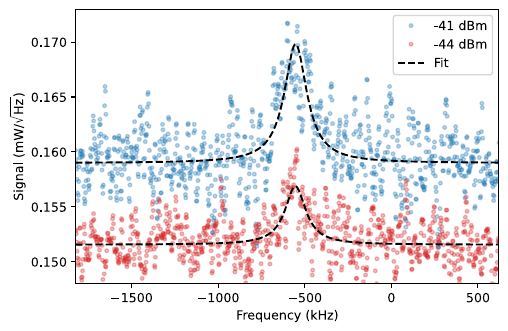}
\caption{\textbf{Calibration of the optomechanical coupling rate.} The amplitude of the scattered peak is compared to a signal of known power (not shown). The area under the peak is calculated from a Lorentzian fit, and the ratio of the fitted peak with the signal of known power yields the optomechanical coupling rate $g_0$.}
\label{FigCalibrationg0}
\end{figure}
The optomechanical coupling $g_0$ is calibrated by sending in a tone at the red sideband, $\omega_\mathrm{c} - \omega_\mathrm{m}$, and comparing the amplitude of the scattered peak with the amplitude of the drive. The resulting signal is shown in Fig.~\ref{FigCalibrationg0}, and we fit a Lorentzian curve to the peak. We extract the area $A$ under the curve, which is proportional to the number of scattered photons, and we compare it to the number of photons in the drive tone, $\hat{a}_\mathrm{sb}^\dagger \hat{a}_\mathrm{sb}$. The optomechanical coupling can be extracted from these amplitudes as 
\begin{equation}
g_0^2 = \frac{A}{|{a}_\mathrm{sb} |^2} \frac{ |1 - \kappa_\mathrm{e}\chi(\omega_\mathrm{c} - \omega_\mathrm{m})|^2 |\chi(\omega_\mathrm{c} - \omega_\mathrm{m})|^2}{|\chi{\omega_\mathrm{c}}|^2 n_\mathrm{m}},
\end{equation}
which contains the ratio of the peak amplitude, as well as the cavity susceptibilities at the red sideband and cavity center (assuming zero detuning), 
\begin{equation}
\begin{aligned}
\chi(\omega_\mathrm{c} - \omega_\mathrm{m}) &= \frac{1}{\kappa/2 + i\omega_\mathrm{m}} \\
\chi(\omega_\mathrm{c}) &= \frac{1}{\kappa/2}.
\end{aligned}
\end{equation}
The thermal mechanical occupation can be calculated as 
\begin{equation}
n_\mathrm{m} = \frac{k_\mathrm{B} T}{\hbar \omega_\mathrm{m}}.
\end{equation}
With these expressions, we fit $g_0 = 2\pi \times 150 \pm 9$~\si{\hertz}.

\subsubsection{Effective linewidth}
We account for optical damping (sideband cooling) by using an effective linewidth, $\gamma_\mathrm{eff}$ that relates to the intrinsic (low-power) mechanical linewidth $\gamma_\mathrm{m}$ via
\begin{equation}
\gamma_\mathrm{eff} = \gamma_\mathrm{m} + \frac{4g_0^2 |\alpha|^2}{\kappa} P = \gamma_\mathrm{m} + \eta P.
\label{Eqeffectivelinewidth}
\end{equation}
Here, $|\alpha|^2$ is the number of photons in the cavity at the red sideband frequency, $\eta$ is some coefficient which we are trying to fit and $P$ is the power sent out from our microwave source at the red sideband frequency. The exact value of $\eta$ depends not only on optomechanical parameters $g_0$ and $\kappa$, but also on the attenuation in the microwave lines between source and sample. However, as long as the latter are constant between measurements, we can use $\eta$ and do not need to know the exact attenuation from our lines. 

To fit the intrinsic linewidth, mechanical frequency and $\eta$, we vary the power sent in on the red sideband. 

\subsubsection{Scale and phonon number calibration}
We do not know a priori the exact scaling between the spectral power that we measure and the expected spectral power of a single photon or phonon. Thus we need to derive an expression for the spectral power in terms of system parameters that can be measured, to use it as a fit function to our measurements. 

Here we use operator formalism as the calibration described in this section is performed close to the quantum regime where vacuum fluctuations should be taken into account. We start from the equations of motion for the cavity field $\hat{a}$ and mechanical ladder operator $\hat{b}$ corresponding to $\tilde{x}$: $\tilde{x} = x_{\rm zpf} (\hat{b}+\hat{b}^\dagger)$. In the sideband-resolved limit, with a drive detuned by $-\omega_\mathrm{m}$ from the cavity frequency (at the red sideband), with $\hat{a}$ separated into a steady state amplitude and some fluctuations $\hat{a} = \alpha + \delta \hat{a}$, the equations of motion can be written in the frequency domain as
\begin{equation}
\begin{aligned}
\hat{a}[\omega] &= \chi_\mathrm{c}[\omega + \Delta]\left(-ig_0 \alpha \left( \hat{b}[\omega] + \hat{b}^\dagger [\omega + 2\omega_\mathrm{m}]\right) + \sqrt{\kappa_\mathrm{i}} \hat{a}_\mathrm{in}[\omega]\right), \\
\hat{b}[\omega] &= \chi_\mathrm{m}[\omega] \left( -ig_0\left(\alpha^* \hat{a}[\omega ] + \alpha \hat{a}^\dagger [\omega + 2\omega_\mathrm{m}] \right) + \sqrt{\gamma_\mathrm{m}} \hat{b}_\mathrm{in}[\omega] \right).
\end{aligned}
\end{equation}
Here, the cavity susceptibility $\chi_\mathrm{c}[\omega] = \frac{1}{\kappa/2 - i\omega}$ and the mechanical susceptibility $\chi_\mathrm{m}[\omega] = \frac{1}{\gamma_\mathrm{m}/2 - i\omega}$ are used as shorthands, and the input fields are $\hat{a}_\mathrm{in}[\omega]$ and $\hat{b}_\mathrm{in}[\omega]$.

In the good cavity limit, we can neglect the terms at $\omega \pm 2\omega_\mathrm{m}$ and $\Delta$. We can replace $\hat{b}[\omega]$ in our expression for $\hat{a}[\omega]$ and compute the spectrum $S_\mathrm{out}$ of the output field $\hat{a}_\mathrm{out}[\omega] = \hat{a}_\mathrm{in}[\omega] - \sqrt{\kappa_\mathrm{e}} \hat{a}[\omega]$. We get
\begin{equation}
S_\mathrm{out} = \frac{1}{2} + \kappa_\mathrm{e}\kappa_\mathrm{i} \left|\tilde{\chi}_\mathrm{c}[\omega]\right|^2 n_\mathrm{cav} + g_0^2 |\alpha|^2 \gamma_\mathrm{m}\kappa_\mathrm{e} \left|\tilde{\chi}_\mathrm{c}[\omega] \chi_\mathrm{m}[\omega]\right|^2 n_\mathrm{th}.
\label{EqSout_goodcavity}
\end{equation}
The three terms come from the external port of the cavity (vacuum), environmental noise at the cavity frequency ($n_\mathrm{cav}$) and mechanical noise (thermal, $n_\mathrm{m}$). The shorthand $\tilde{\chi}_\mathrm{c}[\omega] = \frac{\chi_\mathrm{c}[\omega]}{1 + g_0^2|\alpha|^2 \chi_\mathrm{c}[\omega]\chi_\mathrm{m}[\omega]}$ denotes the cavity susceptibility that is modified due to the optomechanical interaction. In the fitting procedure, we subtract the constant offset, so the factor $1/2$ in Eq.~\eqref{EqSout_goodcavity} disappears. Our final fit has the form $S_\mathrm{fit} = S_c\times(S_\mathrm{out}[\omega] - 1/2)$, and requires knowing $\kappa_\mathrm{e}, \kappa_\mathrm{i}, \gamma_\mathrm{m}, g_0, \alpha, n_\mathrm{cav},$ and $ n_\mathrm{th}$ to extract our fit paramater, scale $S_c$. In favor of knowing $\alpha$ directly, we can use $g_0^2|\alpha|^2 = \frac{\kappa\eta P}{4}$ using the parameter $\eta$ defined in Eq.~\eqref{Eqeffectivelinewidth}.

\begin{figure}
\includegraphics[width = 0.48\textwidth]{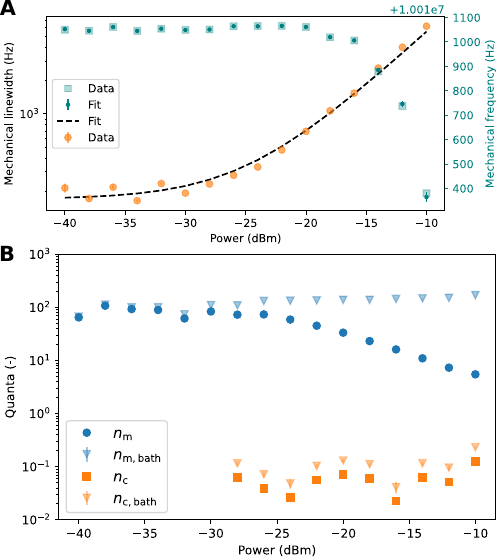}
\caption{\textbf{Power calibration. A}: Mechanical linewidths (orange) and frequencies (teal) for various driving powers on the red sideband. \textbf{B}: Extracted mechanical occupation $n_\mathrm{m}$ and cavity photon number $n_\mathrm{c}$, as well as the equivalent quanta in the respective thermal baths.}
\label{FigSidebandcalibration}
\end{figure}

The fit procedure is as follows: We assume that at sufficiently low red-detuned drive power, the mechanical occupation is unperturbed by the drive. For the points at the lowest powers shown in Fig.~\ref{FigSidebandcalibration}\textbf{A}, below $-32$~dB, $n_\mathrm{m}$ is known from the reference temperature obtained from the thermalization step. Then we use the power-dependence parameter $\eta$ obtained from the effective linewidth step, and fit the data at all powers to obtain $n_\mathrm{m}$ and $n_\mathrm{c}$ at each power. The effective linewidth is shown in Fig.~\ref{FigSidebandcalibration}\textbf{A}, with intrinsic mechanical linewidth $\gamma_\mathrm{m} = 168.9 \pm 9.5$~\si{\hertz} and power factor $\eta = 5.37\cdot 10^4$~\si{\hertz\per\milli\watt} completing the fit.

With our initial guess of the occupation numbers, we subsequently refine the fits of our scale parameter and $\eta$ (now using the whole dataset). Using the updated values, we re-fit $n_\mathrm{m}$ and $n_\mathrm{c}$ at all powers, and the final result is shown in Fig.~\ref{FigSidebandcalibration}\textbf{B}. In the plots, we have used the final $\eta = 5.37\cdot 10^4$~\si{\hertz\per\milli\watt}, and scale $S_c = 2.04\cdot 10^{-15}~\frac{\mathrm{W}/\sqrt{\mathrm{Hz}}}{n_\mathrm{c}}$.

\subsection{Data filtering}\label{SecSIDataFiltering}
\begin{figure}
\includegraphics[width = 0.48\textwidth]{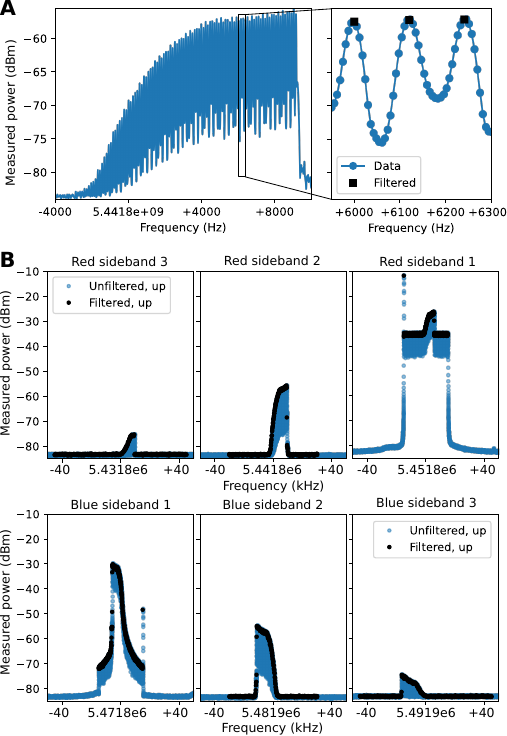}
\caption{\textbf{Data filtering. A}: The measured power on the spectrum analyzer shows periodic peaks at the sideband drive frequency steps. We filter out the peaks located at these frequencies (black squares). \textbf{B}: All sidebands as recorded on the spectrum analyzer (blue) and filtered (black). Clearly visible is the noise pedestal due to the drive at the first red sideband. The sharp peak at the edge of the pedestal is from the vector network analyzer start/stop frequency.}
\label{FigDatafiltering}
\end{figure}
Due to limitations of our vector network analyzer, we drive our system at a series of discrete frequencies instead of a continuous sweep. This way, we can control the step size and direction, which allows us to do a bidirectional measurement where the drive frequency first increases and then decreases. We record the response of our system on a separate spectrum analyzer (SA), which integrates for the full duration of the increasing/decreasing segments separately. The response measured at the second red sideband ($\omega_\mathrm{c} - 2 \omega_\mathrm{m}$) during the upwards frequency sweep is shown in Fig.~\ref{FigDatafiltering}\textbf{A}. 

Our SA records the spectrum using a much finer set of points than the frequencies at which the vector network analyzer outputs its drive. This condition creates a periodic pattern of peaks in the SA-recorded spectrum, which obscures the true shape of the curve. We are only interested in the tops of the peaks, since there is no sideband drive at the frequencies between the peaks. We filter the SA points in a small bandwidth around each of the vector network analyzer output frequencies, and integrate the power over that bandwidth. The filtered signal is a much shorter set of points that traces out the tops of the peaks. We have plotted the unfiltered data and the filtered signals as blue dots and black squares respectivel in Fig.~\ref{FigDatafiltering}\textbf{A}.

\subsection{Exclusion of other sources of nonlinearity}\label{SecSINonlinearities}
\subsubsection{Electrostatic nonlinearity: Average potential offset}
\begin{figure}
\includegraphics[width = 0.5\textwidth]{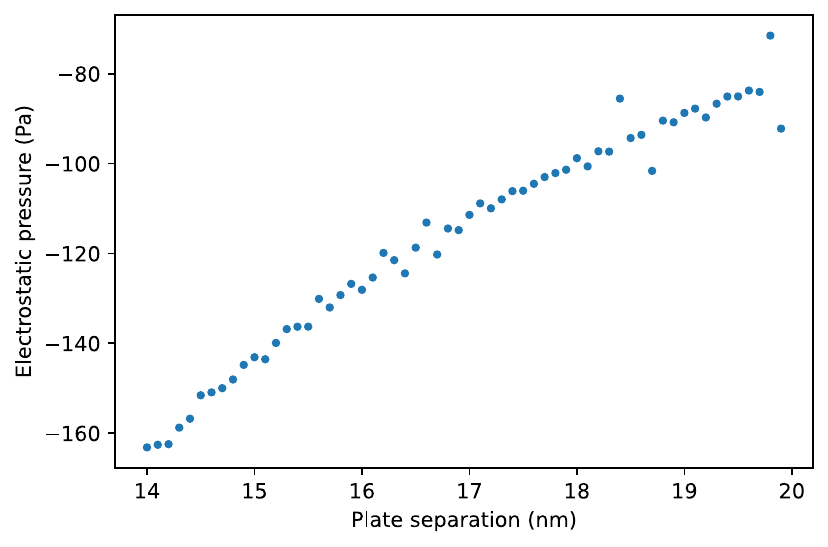}
\caption{\textbf{Electrostatic effects.} Electrostatic pressure due to a \SI{1}{\volt} static potential offset between the drum plates, as a function of their separation distance $d$. Due to the small energies involved and the geometry of the simulation, it is sensitive to mesh-related inaccuracies.}
\label{FigElectrostaticpressure}
\end{figure}

The top and bottom plates of our drum form a capacitance separated by a vacuum gap at equilibrium of $d' \simeq 15$~\si{\nano\meter}. Any static average potential difference between the plates exerts an attractive force between the plates, which has been used in a similar system to tune the vacuum gap~\cite{Andrews2015}. In our system, the top and bottom plates are connected galvanically through the superconducting cavity, thus they are two sides of the same superconductor. In the absence of a large thermal gradient across our device, we expect the average potential difference between the plates to be negligibly small. Nonetheless, if there were a \SI{1}{\volt} static average potential difference, this would correspond to a pressure of $P_\mathrm{elec} \simeq 150$ \si{\pascal} for $d' = 15$~\si{\nano\meter} (based on a COMSOL simulation, as shown in Fig.~\ref{FigElectrostaticpressure}). This pressure is negligible compared to the Casimir pressure, which at this distance is $P_\mathrm{c} = 12.0 \pm 0.6$~\si{\kilo\pascal}. 

Our experiment is not directly sensitive to the absolute value of the (Casimir) pressure, and neither is it directly sensitive to any absolute pressure from the electrostatic force. It is  sensitive to the \emph{nonlinearity} that originates from these effects: The electrostatic force contributes a softening nonlinearity that scales with $P_\mathrm{elec} \propto d^{-2}$. This scaling is much weaker than the Casimir force scaling, $P_\mathrm{c} \propto d^{-3.2}$. Since the magnitude of the electrostatic force is much less and the nonlinearity scales much weaker than the Casimir force, the effect from the average potential difference is negligible. 

\subsubsection{Electrostatic nonlinearity: Potential patches}
It is well known that crystal grain orientations can cause local differences in the electrostatic potential~\cite{Camp1991}, also known as potential patches~\cite{Speake2003}. This means that although two closely spaced conductors may have the same \emph{average} potential, there may still be a non-zero attractive electrostatic force between the conductors. Numerical simulations of randomized patch geometries can be used to estimate the force contributed by these potential patches~\cite{Speake2003,Ke2023,deJong2024}. Such numerical simulations rely on accurate information about the (lateral) sizes of these patches, as well as their voltage distribution~\cite{Behunin2012,Behunin2012a}.

\begin{figure}
\includegraphics[width = 0.5\textwidth]{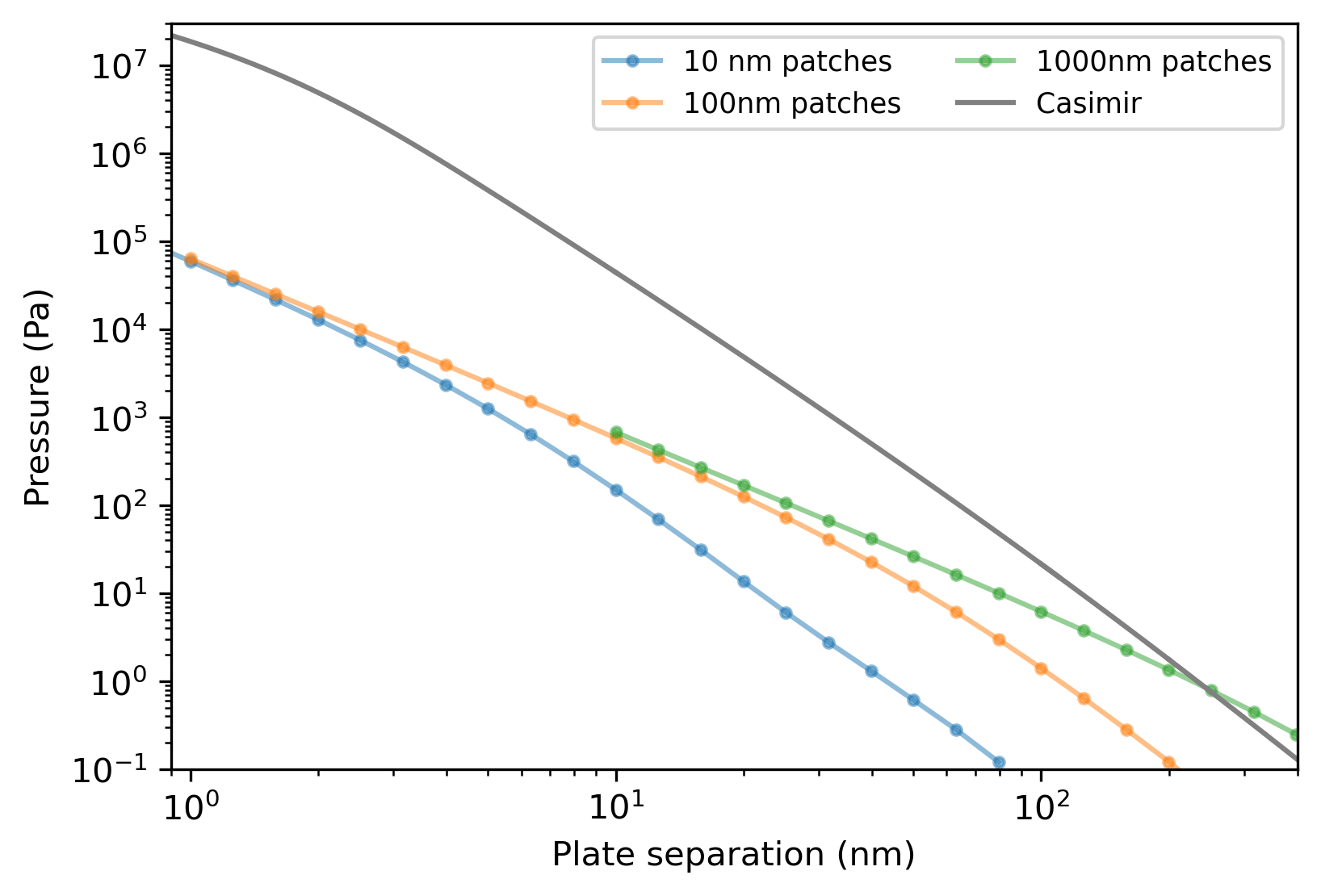}
\caption{\textbf{Patch pressure.} Calculated Casimir pressure and patch potential pressure for patches normally distributed with a standard deviation of \SI{100}{\milli\volt}, for various patch sizes.}
\label{FigCasimirvsPatchpressure}
\end{figure}

We experimentally measure the patches in our aluminium to have a characteristic size of $\ell = 158$~\si{\nano\meter} and a potential intensity that falls well within a normal distribution width $\sigma = 100$~\si{\milli\volt}. With these numbers, we overestimate the pressure exerted by potential patches by about a factor 10. Nonetheless, we find that the patch pressure is several orders of magnitude smaller than the Casimir pressure at our equilibrium position $d' = 15.1$~\si{\nano\meter}, as shown in Fig.~\ref{FigCasimirvsPatchpressure}. We exclude that potential patches are responsible for the nonlinearity that we observe in our mechanical resonator. We describe the experimental details in the next paragraphs.

\begin{figure}
\includegraphics[width = 0.5\textwidth]{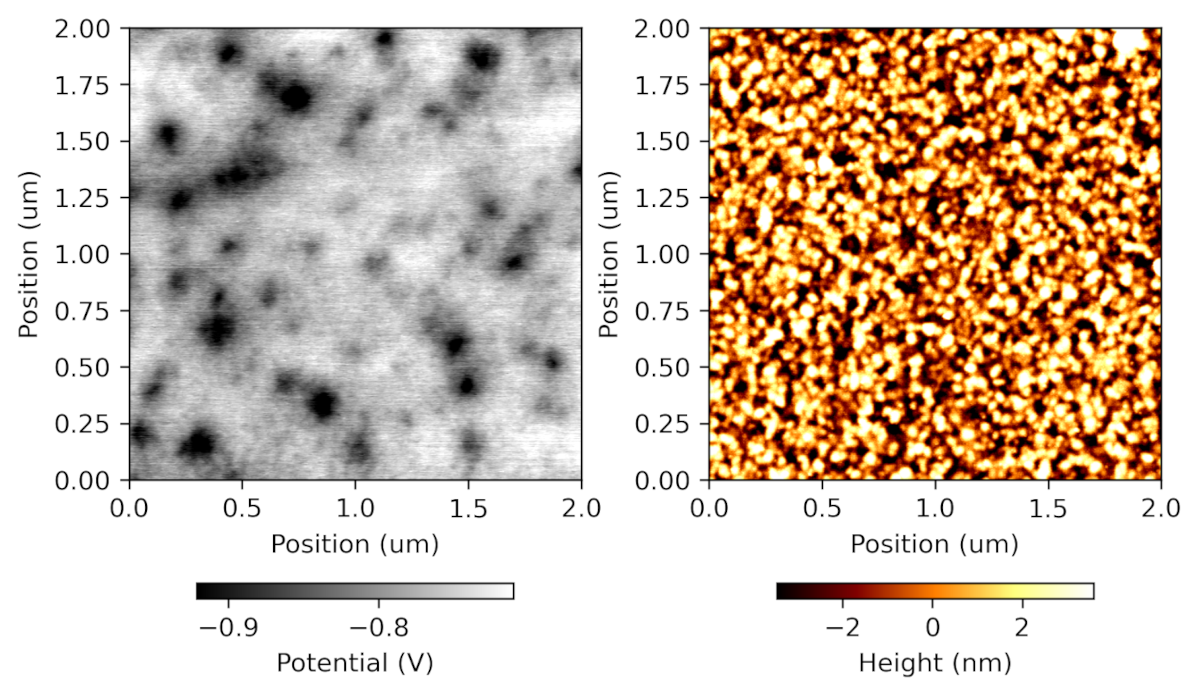}
\caption{\textbf{Potential patches.} Measured surface potential (left) and height (right) of a $2\times2$~\si{\micro\meter\squared} aluminium film from the same fabrication batch as the sample studied in the main text.}
\label{FigPotentialPatches}
\end{figure}

\paragraph*{KPFM measurement of patches}\mbox{}\\
We use Kelvin Probe Force Microscopy (KPFM) to locally measure the potential on our sample, which is a well-established technique based on a conductive atomic force microscope cantilever~\cite{Behunin2014,Garrett2015}. We use a sample that was fabricated in the same batch as the sample reported in the main paper. The KPFM measurement results are plotted in Fig.~\ref{FigPotentialPatches}, where we see characteristic aluminium spots where the potential is $\simeq 100-200$~\si{\milli\volt} below the mean. This corresponds to the work function difference between the different crystalline orientations of aluminium: The work function of aluminium in the $\left\langle 100 \right\rangle$-direction is \SI{4.20}{\eV}, in the $\left\langle 110 \right\rangle$-direction it is \SI{4.06}{\eV} and in the $\left\langle 111 \right\rangle$-direction it is \SI{4.26}{\eV}~\cite{Haynes2015}. There is no correlation between the potential and height, which are measured simultaneously.

\begin{figure}
\includegraphics[width = 0.5\textwidth]{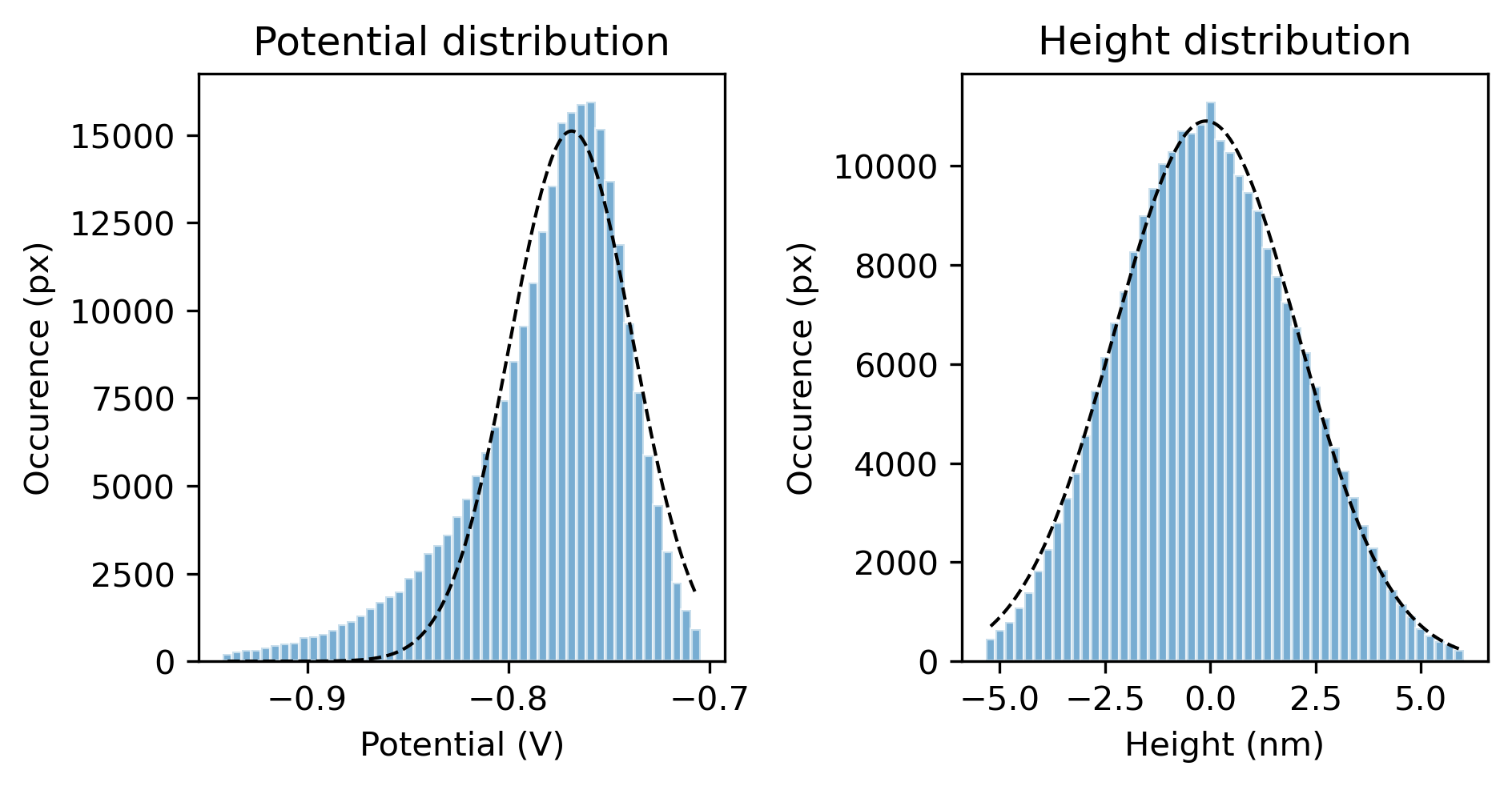}
\caption{\textbf{Potential patch statistics.} Distribution of the measured potentials and heights for the area shown in Fig.~\ref{FigPotentialPatches}, and Gaussian fits (dashed black lines).}
\label{FigPotentialPatches2}
\end{figure}

\paragraph*{Patch intensity}\mbox{}\\
We expect a Gaussian distribution of the potentials, since the crystalline orientations should be random. But when we bin the measurement points of the potential, the Gaussian fit shown in Fig.~\ref{FigPotentialPatches2} is not a good fit. There is a distinct tail towards lower potentials, which represents the spots seen in Fig.~\ref{FigPotentialPatches}. The Gaussian fit is centered around a mean value \SI{-768}{\milli\volt} with a standard deviation $\sigma$ of \SI{30.4}{\milli\volt}. The mean potential seen in Fig.~\ref{FigPotentialPatches} is due to the insulating substrate due to which the sampled area is not well galvanically connected to ground: Our KPFM tip touches down before the measurement and this transfers some charge to the sample. Any excess charge leaks out over time (timescale of approximately 30 minutes in air), so our measurement shows some offset potential. 

\begin{figure}
\includegraphics[width = 0.5\textwidth]{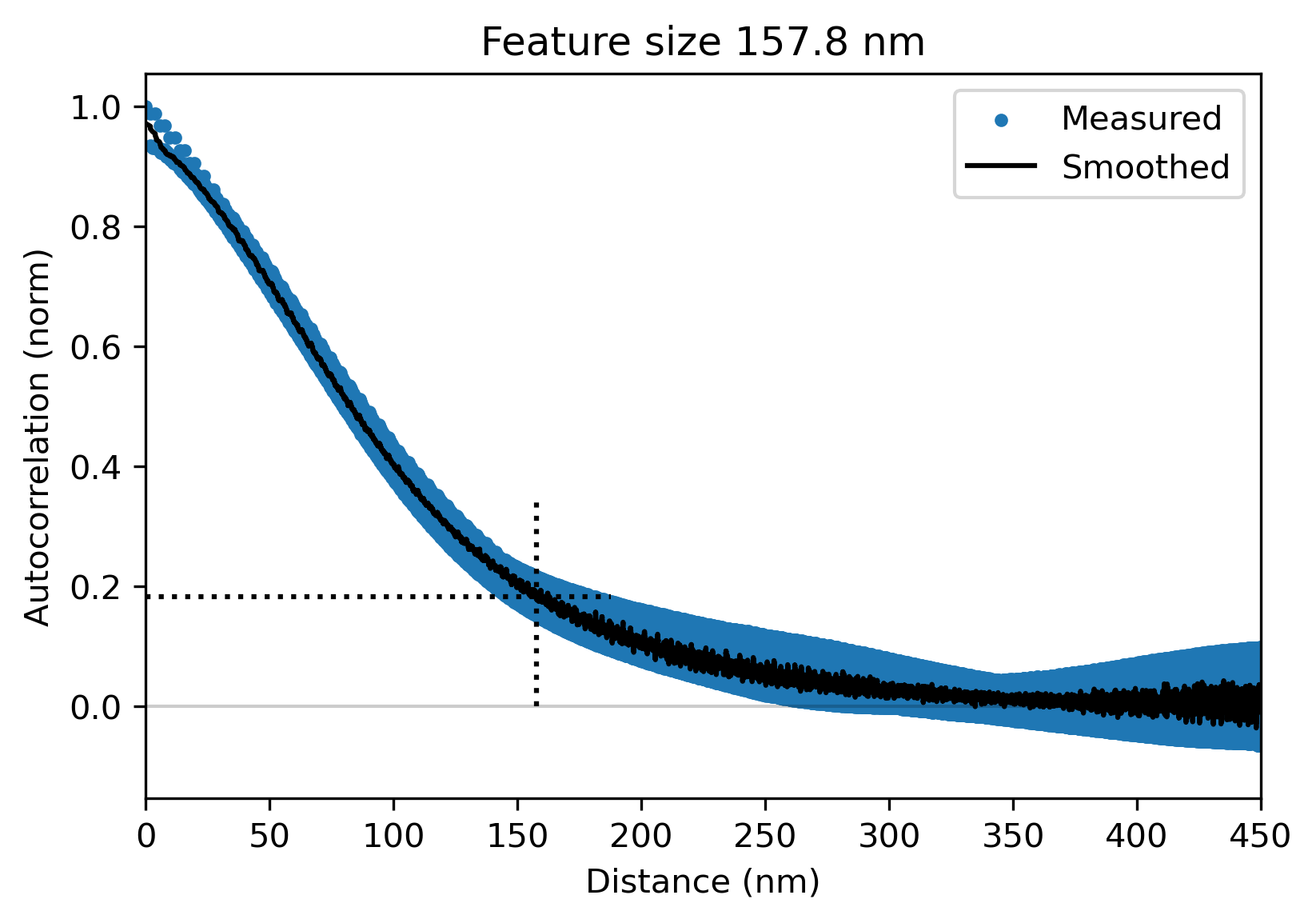}
\caption{\textbf{Potential patch size.} Autocorrelation of the potential shown in Fig.~\ref{FigPotentialPatches}. The average patch size $\ell_\mathrm{p}$ is the distance at which the normalized autocorrelation drops to $1/2e \simeq 0.184$ (dotted lines), which is \SI{157.8}{\nano\meter} for our aluminium film.}
\label{FigPotentialPatches3}
\end{figure}

\paragraph*{Patch size}\mbox{}\\
Besides the intensities of the potential patches, their spatial extent (average patch feature size $\ell$) strongly affects the distance scaling of the electrostatic force they exert~\cite{Ke2023,deJong2024}. Recent simulations show that the lateral resolution of the KPFM probe can lead to an overestimation of the patch size and an underestimation of the patch intensity~\cite{Shi2024}. We use ASYELEC02 probes in our AFM, which have a nominal tip radius of \SI{25}{\nano\meter}, and operate in tapping mode with an amplitude \SI{10}{\nano\meter}. With these numbers, we can estimate the lateral resolution of our KPFM measurement using\cite{Leveque2005} $\ell_\mathrm{min} = \sqrt{RH} \simeq 16$~\si{\nano\meter} where $R$ is the tip radius and $H$ is the minimal tip-sample distance. We determine the average size of the experimentally measured patches using the \verb|correlate2d| function from Scipy, which is plotted in Fig.~\ref{FigPotentialPatches3}. We find the average patch size $\ell_\mathrm{p}$ is \SI{158}{\nano\meter}, which is well above our minimal lateral resolution, $\ell_\mathrm{p} \gg \ell_\mathrm{min}$. Thus there is no reason to assume we significantly overestimate the characteristic patch size $\ell$ or underestimate the patch intensity distribution width $\sigma$.
Our KPFM measurement also measures the height variation in our sample. The height is well-described by a Gaussian fit centered around \SI{-0.1}{\nano\meter} with a $\sigma_\mathrm{h}$ of \SI{2.19}{\nano\meter} and a characteristic feature size $\ell_\mathrm{height} = 32$~\si{\nano\meter}, as shown in Fig.~\ref{FigPotentialPatches2}.

\paragraph*{Patch textures reproducibility}\mbox{}\\
Our patch characterization was performed on a sample that was processed identically (i.e., in the same batch) as the sample in which we measured the Casimir force. We have measured the patch statistics on 8 separate areas on the aluminium parts of the sample, and obtained similar patch intensities and patch sizes. The average patch intensity distribution width from all 8 areas $\sigma_\mathrm{mean} = 36$~\si{\milli\volt}, and the average patch size is $\ell_\mathrm{mean} = 160$~\si{\nano\meter}, with both parameters varying minimally between different measurement areas.

While some processing steps greatly affect the size and intensity of potential patches~\cite{Garrett2020}, we have performed KPFM measurements on multiple samples (from different batches) and found our processing does not significantly affect the potential patches. The patch intensity does not appear affected by our release etch step, as we have measured $\sigma_\mathrm{mean} = 41$~\si{\milli\volt} before release ($\sigma_\mathrm{mean} = 36$~\si{\milli\volt} after). The heat treatment step has the largest effect on the patch intensity, $\sigma_\mathrm{mean} = 63$~\si{\milli\volt} before any heat treatment, $\sigma_\mathrm{mean} = 55$~\si{\milli\volt} after 15 minutes of \SI{200}{\celsius}, $\sigma_\mathrm{mean} = 47$~\si{\milli\volt} after 15 minutes of \SI{240}{\celsius}, $\sigma_\mathrm{mean} = 42$~\si{\milli\volt} after 15 minutes of \SI{260}{\celsius}, and $\sigma_\mathrm{mean} = 41$~\si{\milli\volt} after 15 minutes of \SI{280}{\celsius}. These findings appear consistent with potential patches that originate solely from the crystalline structure of the material. Our patch measurements were performed on a sample that has been processed simultaneously and identically to the experimentally reported sample, so we consider them to be representative of the patches present in the drum we tested.

\paragraph*{Estimate of patch forces}\mbox{}\\
We calculate the effect of patches for our drum geometry using the simulation framework we developed in an earlier work~\cite{deJong2024}. We generate Voronoi patterns based on randomized patches of various sizes, with normally distributed potentials with a $\sigma = 100$~\si{\milli\volt}. Our patterns consist of $400$ patches, and we compute the electrostatic energy. From the variation of energy with plate separation, we extract the pressure from the patch potentials for various plate separations as shown in Fig.~\ref{FigCasimirvsPatchpressure}. We find the well-known limit behaviors with the patch size~\cite{Speake2003}: When $\ell \gg d$, $P_\mathrm{patch} \propto d^{-2}$, while when $\ell \ll d$, $P_\mathrm{patch} \propto d^{-4}$. The regime where $\ell \gg d$ leads to the greatest patch pressure in absolute sense, yet this is still several orders of magnitude smaller than the Casimir pressure for our plate separation $d = 15$~\si{\nano\meter}. We find $P_\mathrm{patch} = 230$~\si{\pascal} at a vacuum gap of $15$~\si{\nano\meter}, much smaller than the Casimir pressure $P_\mathrm{c} = 12000\pm 600$~\si{\pascal}, as shown in Fig.~\ref{FigCasimirvsPatchpressure}. We note that in Casimir experiments at larger distances ($d \gtrsim 200$~\si{\nano\meter}), the Casimir pressure can be comparable in amplitude to the patch pressure.

\subsubsection{Mechanical nonlinearity: Geometric origin}
There is a significant body of literature on the mechanical nonlinearities with a purely geometric origin. In our experimental platform of superconducting Aluminium drum resonators, Ref.~\cite{Cattiaux2020} provides an excellent theoretical background that is experimentally tested in Ref.~\cite{Cattiaux2020a}. The nonlinearity due to geometry is a \emph{hardening} nonlinearity in these drum resonators, both on theoretical grounds and experimental evidence~\cite{Cattiaux2020,Cattiaux2020a}. The Casimir force contributes a \emph{softening} nonlinearity, which is exactly what is described in the main text, and thus we exclude the geometric nonlinearity on qualitative grounds. 

Furthermore, we can exclude the geometric mechanical nonlinearity on quantitative grounds. We follow the method of Ref.~\cite{Cattiaux2020}, which is based on Kirchoff-Love theory for circular membranes. The first assumption in this method is that the resonator is essentially a 2D circular membrane with coordinates $(r,\theta)$, which has modes purely moving in the out-of-plane $z$ direction. From COMSOL simulations, we verify that the mode shown in Fig.~\ref{FigDrumSimulations}\textbf{B} consists for $>99\%$ of motion in the $z$-direction. Thus we separate the variables of the function $f_{n,m}$ describing the motion of our structure for the $(n,m)$ mode,
\begin{equation}
f_{n,m}(r,\theta,t) = z_{n,m}(t)\psi_{n,m}(r,\theta).
\end{equation}
The part $z_{n,m}(t)$ describes the oscillating motion, and $\psi_{n,m}(r,\theta)$ is the normalized mode shape. The mass and stiffness parameters of the fundamental mode (which is the one we study) are given as~\cite{Cattiaux2020}
\begin{equation}
\begin{aligned}
\mathcal{M}_{0,0} &= \rho h \int_0^{2\pi} \int_0^{R_\mathrm{d}} \left( \psi_{0,0}(r,\theta)\right)^2 r dr d\theta, \\
\mathcal{K}_{0,0} &= \frac{1}{12}\frac{E h^3}{1-\nu_\mathrm{r}^2} \int_0^{2\pi} \int_0^{R_\mathrm{d}} \left( \psi_{0,0}(r,\theta) \Delta^2 \psi_{0,0}(r,\theta)\right) r drd\theta \\
&+ h\sigma_0 \int_0^{2\pi} \int_0^{R_\mathrm{d}} \left( \psi_{0,0}(r,\theta) \Delta \psi_{0,0}(r,\theta)\right) r drd\theta.
\label{EqLinearmodeparameters}
\end{aligned}
\end{equation}
We denote the material parameters: $\rho$ is the density of the drum material, $E$ is the Young's modulus, $\nu_\mathrm{r}$ the Poisson's ratio, and $\sigma_0$ the stress (negative for tensile stress). The drum geometry is taken into account via the drum radius $R_\mathrm{d}$ and thickness $h$. Finally, 
\begin{equation}
\Delta ... = \frac{1}{r}\frac{\partial ...}{\partial r}\left( r\frac{\partial ...}{\partial r} \right) + \frac{1}{r^2} \frac{\partial^2 ...}{\partial \theta^2}
\end{equation}
is the Laplacian operator in polar coordinates. From Eq.~\eqref{EqLinearmodeparameters}, we can calculate the mode frequency $\omega_{0,0} = \sqrt{\mathcal{K}_{0,0}/\mathcal{M}_{0,0}}$. 

In the COMSOL model shown in Fig.~\ref{FigDrumSimulations}, we use material parameters $\rho = 2700$~\si{\kilo\gram\per\meter\cubed}, $E=76.6$~\si{\giga\pascal}, $\nu_\mathrm{r} = 0.32$, and the average value of the von Mises stress over the drum domain, $\sigma_0 = -270$~\si{\mega\pascal} (following the convention of Ref.~\cite{Cattiaux2020}, tensile stress is negative in the analytical description, while in the COMSOL plots tensile stress is positive). Combined with the drum diameter $2R_\mathrm{d} = 11.3$~\si{\micro\meter} and the thickness of the evaporated Al $h = 120$~\si{\nano\meter}, we can evaluate Eq.~\eqref{EqLinearmodeparameters}. We find $\mathcal{M}_{0,0} = 1.38\times 10^{-13}$~\si{\kilo\gram \meter\squared} and $\mathcal{K}_{0,0} = 1467$~\si{\newton\meter}. The resulting frequency $\omega_{0,0} = 2\pi \times 16.4$~\si{\mega\hertz} is close to our bare resonator frequency $\omega_\mathrm{r} = 2\pi\times 16.3$~\si{\mega\hertz}.

The geometric nonlinearity can be captured in a cubic term ($\propto x^3$), and its coefficient for the fundamental mode can be found from the expression~\cite{Cattiaux2020}
\begin{equation}
\tilde{\mathcal{K}}_{0,0} = -\frac{Eh}{R_\mathrm{d}^2} \frac{C^{(1)}_{0,0}}{1-\nu_\mathrm{r}} \int_0^{2\pi} \int_0^{R_\mathrm{d}} \left( \psi_{0,0}(r,\theta) \Delta \psi_{0,0}(r,\theta)\right) r drd\theta
\end{equation}
where $C^{(1)}_{0,0} = 0.389664$ as tabulated in Table IV of Ref.~\cite{Cattiaux2020}. From our COMSOL simulations, we obtain $\tilde{\mathcal{K}}_{0,0} = 3.55\times 10^{15}$~\si{\newton\per\meter}. We then relate this to the coefficient $\alpha_\mathrm{D}$ of the Duffing term,
\begin{equation}
\ddot{x} + \gamma_\mathrm{r}\dot{x} +\omega_\mathrm{r} x +\alpha_\mathrm{D} x^3 = \frac{F_\mathrm{d}}{m_\mathrm{eff}},
\end{equation}
where $\alpha_\mathrm{D} = \tilde{\mathcal{K}}_{0,0} / \mathcal{M}_{0,0} = +2.57\times 10^{28}$~\si{\per\meter\squared\per\second\squared}. This value is close to the experimental fit of Refs.~\cite{Cattiaux2020,Cattiaux2020a}, which is $\alpha_\mathrm{D} = +7\times 10^{27}$~\si{\per\meter\squared\per\second\squared} in a drum resonator that is of nearly identical design to the one in this work.

From the coefficient $\alpha_\mathrm{D}$, we can estimate that a $+1$~\si{\hertz} frequency shift should occur if the motional amplitude is $x \simeq 0.6$~\si{\nano\meter}. At those amplitudes, we see a multiple-kHz \emph{negative} frequency shift corresponding to the Casimir force. Thus we can exclude the geometrical mechanical nonlinearity based on the magnitude and sign of the frequency shift. \\

\subsubsection{Higher-order optomechanical couplings}
The optomechanical cavity is formed by a plate capacitor where one of the plates is mechanically compliant. 
The capacitance is not a linear function of the separation distance $d'$, and $g_0$ only represents the first-order Taylor series of the capacitance in $x/d'$. To estimate the 'higher order' optomechanical couplings that stem from the nonlinearity of the capacitance expansion, we follow the method of~\cite{Cattiaux2020a} derived for the same type of system. The cavity frequency and couplings up to third order in $x$ are
\begin{equation}
\begin{aligned}
\omega_\mathrm{c}(x) &= \omega_\mathrm{c}(0) - \left[ g_0 \frac{x}{x_\mathrm{zpf}} + \frac{g_1}{2} \left( \frac{x}{x_\mathrm{zpf}}\right)^2 + \frac{g_2}{2} \left( \frac{x}{x_\mathrm{zpf}}\right)^3 \right] \\
g_1 &= g_0 \left[ 2 \frac{x_\mathrm{zpf}}{d'} -  3\frac{g_0}{\omega_\mathrm{c}(0)}\right] \\
g_2 &= g_0 \left[ 2 \left(\frac{x_\mathrm{zpf}}{d'} \right)^2 -6 \frac{x_\mathrm{zpf}}{d'} \frac{g_0}{\omega_\mathrm{c}(0)} + 5 \left( \frac{g_0}{\omega_\mathrm{c}(0)}\right)^2 \right]
\end{aligned}
\end{equation}
For a motion amplitude of $x = 0.2$~\si{\nano\meter}, close to the maximum amplitude observed here, the frequency shift from the $g_0$ term is $\Delta\omega_\mathrm{c} \simeq -6.5$~\si{\mega\hertz}, the additional shift from $g_1$ would be $\Delta\omega_\mathrm{c} = -7.5\times 10^{-2}$~\si{\mega\hertz} and the additional shift from $g_2$ would be $\Delta\omega_\mathrm{c} = -8.7\times 10^{-4}$~\si{\mega\hertz}. These higher-order contributions are only a relevant contribution to the frequency shift for extremely large displacements ($\simeq$~\si{\nano\meter}). At these displacements, the frequency shift is much larger than the cavity linewidth, $\Delta\omega_\mathrm{c} \gg \kappa$, so the drive efficiency that we describe in the main text is the most important optomechanical effect to take into account. We emphasize that the nonlinearity from the plate capacitor expansion only affects the \emph{cavity} frequency shift and not the mechanical frequency shift from the Casimir force.

\subsection{Measurements on a second device}
We have repeated the experiments of the main text in a second device, fabricated using an identical design but in a different fabrication run. The parameters of the device were characterized using the same method as for the device in the main text, and they are reported in Table \ref{Tableparameters2}. 

\begin{table}[h]
\renewcommand{\tablename}{\textbf{Table}}
\begin{tabular}{l|l|l}
Quantity & Symbol & Value \\
\hline
Unperturbed frequency & $\omega_\mathrm{r}$ & $2\pi \times 16.710$~\si{\mega\hertz} \\
Mechanical frequency & $\omega_\mathrm{m}$ & $2\pi \times 13.688$~\si{\mega\hertz} \\
Mechanical linewidth & $\gamma_\mathrm{r} \simeq\gamma_\mathrm{m}$ & $2\pi \times 231 \pm 4$~\si{\hertz} \\
Cavity frequency & $\omega_\mathrm{c}$ & $2\pi \times 7.589718$~\si{\giga\hertz} \\
External linewidth & $\kappa_\mathrm{e}$ & $2\pi \times 304.7 \pm 2.8$~\si{\kilo\hertz} \\
Internal linewidth & $\kappa_\mathrm{i}$ & $2\pi \times 608.7 \pm 6.2$~\si{\kilo\hertz} \\
Optomechanical coupling & $g_0$ & $2\pi \times 67 \pm 3$~\si{\hertz} \\
Effective mass & $m_\mathrm{eff}$ & $3.96 \times 10^{-14}$~\si{\kilo\gram} \\
Rest separation without Cas. f. & $d$ & $19.1 \pm 0.5$~\si{\nano\meter} \\
Rest separation & $d'$ & $17.3 \pm 0.5$~\si{\nano\meter} \\
Bottom plate diameter & $2r$ & $11.3$~\si{\micro\meter} \\
Casimir amplitude & $P$ & $(1275 \pm 7) \cdot 10^{-24}~\mathrm{Pa}~\mathrm{m}^{n}$\\
Casimir force scaling exponent & $n$ & $3.193$ \\
\end{tabular}
\caption{\textbf{Parameters of the second device.}}
\label{Tableparameters2}
\end{table}

In the absence of the Casimir force, the devices have a similar mechanical resonance frequency. However, this second device has a slightly different separation of the plates at mechanical zero ($d$), the final mechanical frequency is rather different due to the nonlinear nature of the Casimir force. Due to the lower optomechanical coupling rate, and higher cavity linewidth, it requires more drive power to reach larger displacement amplitudes, and we are limited by the maximum power that our measurement chain can handle. These measurements were performed in a different dilution refrigerator to the experiments in the main text, with a separate set of measurement equipment.

\begin{figure}
\includegraphics[width = 0.48\textwidth]{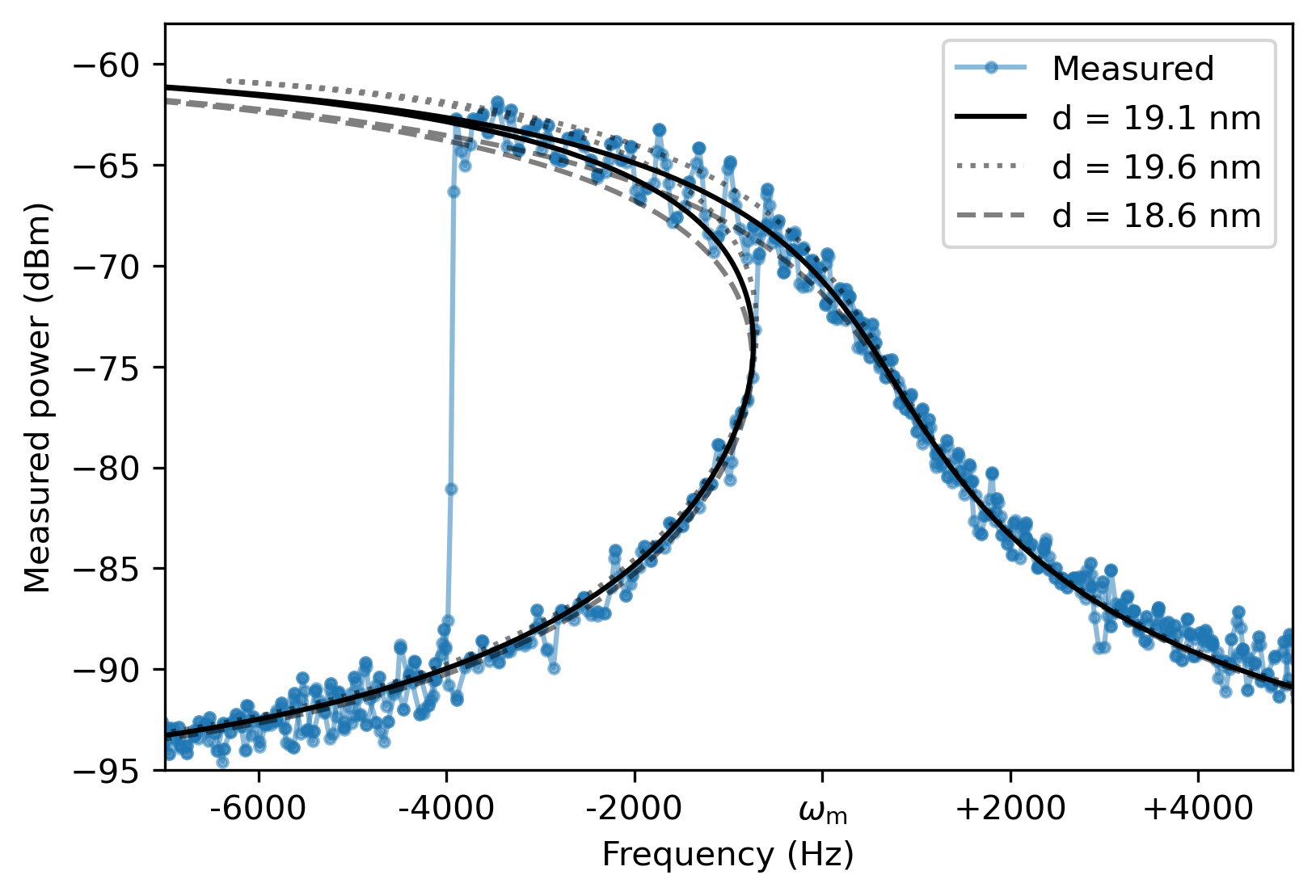}
\caption{\textbf{Measurement of the nonlinearity in a second device.} The measured response (blue) for $10~\mathrm{dBm}$ cavity drive power and $0~\mathrm{dBm}$ red sideband drive power. The theory curves based on the Casimir force for a separation of mechanical zero of $d = 19.1$~\si{\nano\meter} (solid black line), $d= 19.6$~\si{\nano\meter} (dotted black line) and $d = 18.6$~\si{\nano\meter} (dashed black line) are overlaid on the measured data. The $d = 19.1$~\si{\nano\meter} curve is closest to the center of the measured data.}
\label{FigSeconddevice}
\end{figure}

One of the mechanical response curves of the second device is shown in Fig.~\ref{FigSeconddevice}. We have performed the same calibration procedure as in the device of the main text, and extract $d = 19.1 \pm 0.5$~\si{\nano\meter} (solid black line) from a fit to the data. For completeness, we also show the theory curves for $d = 19.6$~\si{\nano\meter} (dotted black line) and $d = 18.6$~\si{\nano\meter} (dashed black line).

\end{document}